\documentclass[usenatbib,usegraphicx]{mn2e}
\usepackage{amsmath}
\usepackage{amssymb}
\usepackage[usenames,dvips]{color}

\newcommand{\rsun}{{R}_{\odot}}
\newcommand{\msun}{{M}_{\odot}}
\newcommand{\lsun}{{L}_{\odot}}

\newcommand{\tcont}{t_{\mathrm{contact}}}

\newcommand{\mtot}{M_\mathrm{tot}}
\newcommand{\Mtot}{\mtot}
\newcommand{\Mtwi}{M_{2,i}}
\newcommand{\Moi}{M_{1,i}}
\newcommand{\Rtwi}{R_{2,i}}
\newcommand{\MR}{$M$-$R$ }

\newcommand{\Mdot}{\dot{M}}
\newcommand{\mdot}{\dot{M}}
\newcommand{\mtwodot}{\dot{M}_2}

\newcommand{\Jdot}{\dot{J}}
\newcommand{\JdotJ}{\frac{\dot{J}}{J}}

\newcommand{\Porb}{P_{\mathrm{orb}}}
\newcommand{\Porbi}{P_{\mathrm{orb},i}}
\newcommand{\Pdot}{\dot{P}_{\mathrm{orb}}}

\newcommand{\RL}{R_L}
\newcommand{\DelR}{\Delta R}

\newcommand{\nrl}{\xi_{R_L}}
\newcommand{\nrtw}{\xi_{R_2}}

\newcommand{\tauth}{\tau_{\mathrm{th}}}
\newcommand{\taum}{\tau_{\mathrm{m}}}

\newcommand{\psic}{\psi_c}
\newcommand{\psici}{\psi_{c,i}}
\newcommand{\rhoc}{\rho_c}
\newcommand{\rhophot}{\rho_{\mathrm{phot}}}
\newcommand{\del}{\nabla}
\newcommand{\delad}{\nabla_{\mathrm{ad}}}
\newcommand{\delrad}{\nabla_{\mathrm{rad}}}
\newcommand{\Gamone}{\Gamma_1}

\newcommand{\Teff}{T_{\mathrm{eff}}}
\newcommand{\Tphot}{T_{\mathrm{phot}}}

\newcommand{\Pphot}{P_{\mathrm{phot}}}
\newcommand{\Tirr}{T_{\mathrm{irr}}}
\newcommand{\Lacc}{L_{\mathrm{acc}}}
\newcommand{\Lsurf}{L_{\mathrm{surf}}}

\newcommand{\plotone}[1]{%
 \includegraphics[width=84mm]{#1}%
}%
\newcommand{\aj}{{AJ}}%
\newcommand{\araa}{{ARA\&A}}%
\newcommand{\apj}{{ApJ}}%
\newcommand{\apjl}{{ApJ}}%
\newcommand{\aap}{{A\&A}}%
\newcommand{\aaps}{{A\&AS}}%
\newcommand{\mnras}{{MNRAS}}%
\newcommand{\pre}{{Phys.~Rev.~E}}%
\newcommand{\pasp}{{PASP}}%
\newcommand{\pasj}{{PASJ}}%
\newcommand{\ee}[2]{\ensuremath{#1 \times 10^{#2}}}

\newcommand{\der}[2]{\ensuremath{\frac{d \,#1}{d#2}}}

\newcommand{\pder}[2]{\ensuremath{\frac{\partial \,#1}{\partial#2}}}

\newcommand{\tder}[3]{\ensuremath{\left(\frac{\partial \,#1}{\partial#2}\right)_{#3}}}

\title{The Thermal Evolution of the Donors in AM CVn Binaries} 
\author[Deloye et al.]{Christopher~J.~Deloye,$^1$\thanks{E-mail:cjdeloye@northwestern.edu} Ronald~E.~Taam,$^1$ Christophe Winisdoerffer,$^2$ and Gilles Chabrier$^2$ \\
$^1$Department of Physics \& Astronomy, Northwestern University, 2131 Tech Drive, Evanston, IL 60208 \\
$^2$Ecole normale sup\'{e}rieure de Lyon, CRAL (UMR CNRS No. 5574), 69364 Lyon Cedex 07, France}

\begin{document}

\maketitle

\begin{abstract}
We calculate the full stellar-structural evolution of donors in AM CVn systems formed through the WD channel coupled to the binary's evolution.  Contrary to assumptions made in prior modelling, these donors are not fully convective over much of the AM CVn phase and do not evolve adiabatically under mass loss indefinitely.  Instead, we identify three distinct phases of evolution: a mass transfer turn-on phase (during which $\Porb$ continues to decrease even after contact, the donor contracts, and the mass transfer rate accelerates to its maximum), a phase in which the donor expands adiabatically in response to mass loss, and a cooling phase beginning at $\Porb \approx$ 45--55 minutes during which the donor contracts. The physics that determines the behaviour in the first and third phases, both of which are new outcomes of this study, are discussed in some detail. We find the overall duration of the turn-on phase to be between $\sim 10^4$-$10^6$ yrs, significantly longer than prior estimates. We predict the donor's luminosity, $L$, and effective temperature, $\Teff$. During the adiabatic expansion phase (ignoring irradiation effects), $L\approx10^{-6}$--$10^{-4} \lsun$ and $\Teff\approx1000$--$1800$ K.  However, the flux generated in the accretion flow dominates the donor's intrinsic light at all times. The impact of irradiation on the donor extends the phase of adiabatic expansion to longer $\Porb$, slows the contraction during the cooling phase, and alters the donor's observational characteristics. Irradiated donors during the adiabatic phase can attain surface luminosities up to $\approx10^{-2} \lsun$. We argue that the turn-on and cooling phases both will leave significant imprints on the AM CVn population's $\Porb$-distribution. Finally, we show that the eclipsing AM CVn system SDSS J0926+3624 provides evidence that WD-channel systems with non-zero entropy donors contribute to the AM CVn population, and we discuss the observational signature of the donor in this system.
\end{abstract}

\section{Introduction}
 \label{sec:intro}
The AM Canum Venaticorum (AM CVn) variables are a class of He-rich objects with variability periods of $\approx$5--66 minutes.  Various lines of evidence \citep[see \S 1 of][for a brief summary]{deloye05} indicate these periods are orbital in origin and that ongoing mass-transfer from a low-mass, essentially pure-He donor onto a white dwarf (WD) accretor is taking place.  Thus, AM CVn systems form the WD-accreting class of so-called ultracompact binaries: interacting stellar binaries with orbital periods, $\Porb$, below the minimum $\Porb$ attainable by H-dominated objects \citep[$\approx$70--80 minutes, see, e.g.,][]{kolb99}.  At such short $\Porb$, the binary evolution is driven by orbital angular momentum losses due to gravity-wave (GW) emission.

The AM CVn objects represent an extreme end-product of stellar-binary evolution and at least some of them are examples of WD-WD binaries that survived the transition from a detached phase of GW-driven in-spiral to their current state of stable mass-transfer. There are significant uncertainties concerning aspects of both the prior binary evolution \citep[most importantly the outcome of common-envelope events;][]{paczynski76} and the outcomes of WD-WD binaries initiating mass transfer.  The characteristics of the AM CVn population can thus provide insights into the physics important to both these phases of their prior evolution. It is expected that space-based GW interferometers, such as the proposed \textit{LISA} mission\footnote{http://lisa.nasa.gov/}, will detect essentially the entire galactic population of AM CVns with $\Porb \lesssim 20$ minutes \citep{nelemans01c,deloye05}, providing an observational picture of unprecedented detail.  Future prospects for strong constraints from the AM CVn population on our theories of binary formation and evolution are, thus, very bright.  In the meantime, advances from electromagnetic observations and developing theory will begin addressing these same questions. 

There have been three distinct binary evolution channels proposed to form AM CVn systems. Two of these formation channels involve a series of two common-envelope (CE) events which bring the remnant cores of a main-sequence (MS) binary close enough for GW-emission to drive them into contact, initiating the AM CVn phase.  These channels are distinguished by the state of the proto-donor prior to contact \citep{nelemans01a}.  In what we'll call the He-star channel, the proto-donor is able to ignite He prior to contact. In the other, which we'll refer to as the WD channel, the donor is partially-to-very degenerate \citep{deloye05} and has not undergone any He burning.  The third channel involves an evolved MS star with a WD companion starting stable mass transfer just before it evolves up the red-giant branch \citep{podsiadlowski03}. The star's core is dominated by He, allowing the binary to evolve to ultracompact configurations.  We'll refer to this third channel as the evolved-MS channel.

In mass transferring binaries, the donor's structural response to mass loss plays a central role in determining the binary's evolution \citep[see, e.g.,][]{deloye03}. The extent to which the donors from  each formation channel have been modelled varies considerably.  He-star channel donors have only been modelled by a fit \citep{nelemans01a} to a single relevant binary evolution calculation by \citet{tutukov89}.  The WD channel donors have been considered more extensively.  \citet{nelemans01a} modelled these donors as fully-degenerate WDs obeying the zero-temperature He WD mass-radius (\MR) relation.  Later, \citet{deloye05}, using donor structural models developed for ultracompact low-mass X-ray binaries \citep{deloye03}, modelled the evolution of non-zero entropy donors from the WD-channel.  \citet{podsiadlowski03} has presented a detailed study of the evolved-MS star channel.

In this paper, we focus on the WD channel systems.  The \citet{deloye05} study had several limitations: they assumed the donors to be fully convective and to evolve adiabatically under mass loss due to the very large mass transfer rates, $\Mdot$,  produced in AM CVns.  With these assumptions, the evolution is completely determined by the total mass of the two components and the donor's specific entropy, which determines the \MR relation the donor follows and, amongst other system parameters, the binary's $\Mdot(\Porb)$ evolution.  The inclusion of non-zero entropy donors in this population resulted in shifting systems to larger $\Mdot$ at fixed $\Porb$ and allowed systems to evolve out to longer $\Porb$ as compared to a population of zero-temperature donors.  This provided an observational diagnostic of an AM CVn system's initial conditions.  The \citet{deloye05} calculation also indicated that WD channel systems with hot donors and He-star channel systems would not be distinguishable based on $\Mdot$ determinations alone.

Since the entropy structure of the donor determines its response to mass loss (see the discussion in \S \ref{sec:donorev_preR2min} and Appendix \ref{app:R_response}), assuming a fully convective structure will overestimate the binary's $\Porb$ evolution rate if the donor's true structure in not adiabatic.  Further, while assuming the donor responds adiabatically to mass-loss is certainly valid at early stages of the AM CVn phase when $\Mdot$ is very large, this assumption may break down as the binary evolves outward in $\Porb$ to lower $\Mdot$. 

To rectify these shortcomings and provide  predictions for the donor's luminosity and effective temperature,  we undertook a study to model the donors within the context of a full stellar structural calculation coupled to the evolution of the AM CVn binary.  In particular, we sought to determine the donor's observational signatures as a function of initial system parameters and $\Porb$, as well as to determine how a complete treatment of the donor's structure alters the conclusions of earlier studies.  We also sought to model accurately the critical phase of mass transfer turn-on, when the donor is coming into contact and a transition in the $\Porb$-evolution from GW-dominated in-spiral to the mass-transfer dominated expansion occurs. This phase of WD channel system evolution has not been modelled in any detail prior to this study.

To carry out the computations, we developed a new stellar evolution code to perform the coupled donor/binary calculations.  A  brief description of this code,  detailing our numerical modelling of the donor and binary is provided in \S \ref{sec:calcdetails}.  In \S \ref{sec:initial_models}, we describe the procedure for determining the range of initial donor degeneracy and donor structural models used for our AM CVn phase calculations. We present the results of our calculations in \S \ref{sec:ev_general} and discuss these results and several of their applications in \S \ref{sec:discussion}. Finally, we summarize in \S \ref{sec:summary}.

 \section{Details of Our Numeric Calculations}
 \label{sec:calcdetails}
To achieve a necessary level of flexibility in terms of input physics and 
defining system of equations for this and related projects, we implemented 
a stellar evolution code capable of handling most 1D stellar 
evolution calculations expressed as boundary-value problems.  This code is 
written in C++ and is comprised of a suite of abstract classes defining the 
necessary components of a stellar evolution calculation.  At its core
is a generalized 1D relaxation-method boundary-value problem (BVP) 
solver \citep[see, e.g, \S 17.3 of][]{press92} that utilizes the sparse-matrix 
solver UMFPACK\footnote{http://www.cise.ufl.edu/research/sparse/umfpack/} 
to perform the necessary matrix inversions.  The defining system of ordinary 
differential equations (ODEs) and necessary input physics are implemented as separate abstract 
classes, providing the desired flexibility. Utilizing this new code, we 
solve for both the donor's structure and binary parameters within a single 
set of relaxation iterations at each time step.  Below we detail our numerical 
treatment for both components of our system and describe the input physics.

\subsection{Calculation of the Donor's Structure}
\label{sec:calcdetails_donor}
For the donor, we solve the standard set of 1D hydrostatic stellar structure 
equations augmented by an automatic mesh allocation algorithm \citep{eggleton71} 
on a 200-point non-lagrangian mesh: 
\begin{gather}
\der{\ln P}{m} = -\frac{G m^2}{4 \pi r^4 P}\,, \label{eq:dPdm}\\
\der{\ln T}{m} = \del \,\der{\ln P}{m}\,,\\
\der{\ln r}{m} = \frac{1}{4 \pi r^3 \rho}\,,\label{eq:drdm}\\
\der{l}{m}     = \epsilon - c_P \left( \pder{T}{t} - \delad \frac{T}{P} \pder{P}{t} \right)\,,\label{eq:dldm}\\
\der{q}{m}     = \frac{1}{\theta} \left( -\alpha_1 \frac{\ln P_c}{\ln P_e} \der{\ln P}{m} + \frac{\alpha_2}{m} \right)\,. \label{eq:dqdm}
\end{gather}
Here $\rho$, $T$, $P$, $r$, and $l$, are the density, temperature, pressure, 
radius, and luminosity at each mesh point, $m$ is the mass interior to each 
mesh point,  $t$ is time, $\epsilon$ is the local nuclear energy generation 
rate (which we set to zero throughout this work), and  $c_P$ is the specific 
heat at constant pressure. In equation (\ref{eq:dqdm})---which determines the 
mesh allocation---$P_c$ and $P_e$ are the pressures at the stellar centre and 
exterior point, respectively, $\alpha_1,\, \alpha_2$ are constants that weight 
the importance of pressure and mass gradients in determining the mesh spacing, 
and $\theta$ is an overall normalization constant determined implicitly at each 
time step. The quantity $\delad \equiv (d \ln T/d \ln P)_s$, where $s$ is the 
specific entropy, and  $\del \equiv d \ln T/d \ln P$ describes the actual run 
of $T$ with $P$ within the donor.  We take $\del = \delrad$ when 
$\delrad < \delad$, where
\begin{equation}
\delrad \equiv \frac{3}{16 \pi a c G}\frac{P \kappa}{T^4}\frac{l}{m}\,,
\end{equation} 
and $\kappa$ is the Rosseland mean opacity.  Otherwise we determine $\del$ using mixing length theory with the mixing length set equal to one pressure scale height. 

The central boundary conditions are determined by the standard central expansions 
of the stellar structure equations.  At the outer boundary,  we equate the 
values of the iterated quantities at 
the outermost mesh point, $T_{e},\,P_{e},\,r_{e},\, l_{e}$, and $m_{e}$ with 
the corresponding values at the photosphere (i.e., we do not calculate a separate atmosphere model).  
Grey atmospheres and the Eddington approximation are assumed.  The photospheric pressure is therefore given by:
\begin{equation}
P_{\mathrm{phot}} = \frac{2}{3} \frac{G M_2}{R_2^2 \kappa_{\mathrm{phot}}}, \label{eq:photosphere}
\end{equation}
where $\kappa_{\mathrm{phot}} = \kappa(\rho_{\mathrm{phot}},\,\Teff)$, $\rhophot$ 
is the photospheric density, and $\Teff$ is the donor's effective temperature. 
This gives the matching conditions $T_{e} = \Teff$, $P_{e}= P_{\mathrm{phot}}$, $r_{e} = R_2$, 
and $m_{e} = M_2$, and $l_{e} = L$. Here $M_2$, $R_2$ and $L$ are the donor's 
total mass, radius, and luminosity, with  $L = 4 \pi \sigma R_2^2 \Teff^4$.  

To produce a smoothly varying outer boundary condition, equation (\ref{eq:photosphere}) is not solved directly each time photospheric values are required. Instead, a $3\times3 $ grid of $(\Teff, g = G M_2/R_2^2 )$ points is constructed that straddle the donor's current photospheric conditions.  We calculate  $P_{\mathrm{phot}}$ from equation (\ref{eq:photosphere}) at these values and use these to determine $P_{\mathrm{phot}}$ by cubic-spline 
interpolations during the iterations.  If a point outside of the current grid is needed, the grid is re-centred and new $P_{\mathrm{phot}}$ values calculated.  The $\Teff$ and $g$ grids have spacings $\Delta \log(\Teff/\mathrm{K}) = 0.05$, $\Delta \log(g/\mathrm{cm\, s^{-2}}) = 0.05$.

\subsection{Calculation of the Binary Parameters}
\label{sec:calcdetails_binary}
To specify the binary system, we add the following three equations to the above set of ODEs:
   \begin{gather}
\der{\ln M_2}{t} = \frac{\mdot_2}{M_2} =  - \frac{\mdot}{M_2} \,,  \label{eq:dm2dt}\\ 
\der{\ln M_1}{t} = \frac{\mdot_1}{M_1} \label{eq:dm1dt}\,,
\end{gather}
and
\begin{multline}
\der{\ln a}{t}  = 2 \frac{\dot{J}}{J} - \frac{1}{M_1+M_2}\\ \times \left[ \frac{\mdot_1}{M_1} \left(M_1 + 2 M_2 \right) +  \frac{\mdot_2} {M_2} \left(M_2 + 2 M_1 \right) \right] \label{eq:dadt} \,,
\end{multline}
where $M_1$ is the accretors mass, $a$ is the orbital separation, $J$ is the orbital angular momentum. The quantities $\mdot_1$, $\mdot_2$ specify the mass evolution rate of the accretor and donor; both are, in general, functions of the binary's and, possibly, the  donor's state. We express the $a$ evolution in the form of equation (\ref{eq:dadt}) to allow for this generality. \emph{However, here, we consider conservative mass transfer, setting $\mdot_1 = \mdot = - \mdot_2$ always}. We consider only gravity wave emission  \citep{landau71} contributions to the $J$ evolution, ignoring the possibility of accretor spin-up in the $J$ evolution.  This latter effect may play a significant role in 
determining whether an AM CVn binary survives the onset of contact 
\citep{nelemans01a, marsh04}. This omission is justified since we are interested 
in understanding the parameter space potentially allowed to AM CVn systems, not 
in a detailed characterization of the actual, realized population.  This latter 
question will be the subject of future studies.

To determine $\mdot_2$ we utilize the prescriptions of \cite{ritter88} and \citet{kolb90}. Specifically, when the donor's Roche radius, $\RL$,  here calculated as \citep{paczynski67}
\begin{equation}
\RL \approx 0.46 a \left(\frac{M_2}{M_1+M_2}\right)^{1/3}\,,
\end{equation}
(valid for $q \equiv M_2/M_1 \leq 0.8$) is larger than $R_2$, the mass loss is modelled as an isothermal flow of a classical gas with a rate
\begin{equation}
-\mdot_2 = \mdot_0 \exp\left(\frac{\DelR}{H_P}\right)\,
\label{eq:m2dot_thin}
\end{equation} 
where $\mdot_0 > 0$ is the mass transfer rate off the donor when $R_2 = \RL$ and
depends on the surface properties \citep{ritter88}, $H_P$ is the pressure scale height at the photosphere, and $\DelR  = R_2 - \RL$.  
When $\RL< R_2$, we model the sub-photospheric contributions with an adiabatic flow \citep{kolb90}.  To do so, we rewrite equation (A17) of \citet{kolb90} in a more general form to allow for degeneracy or non-ideality:   
\begin{equation}
-\mdot_2  = \mdot_0 + 2 \pi F_1(1/q) \frac{\RL^3}{G M_2} \int_{(R_2-\DelR)}^{R_2} F_2(\Gamone) \frac{G m (P \rho)^{1/2}}{r^2} dr \, 
\label{eq:m2dot_thick} 
\end{equation}
where $F_1$ and $F_2$ are defined in \citet{kolb90} and $\Gamone = ( \partial \ln P/\partial  \ln \rho)_{s}$. We determine $\mdot$ implicitly at each timestep, a method that greatly improves the numerical stability \citep{buning06}.

\subsection{Input Physics}
 \label{sec:calcdetails_input_physics}
For the donor's composition, we assume a fixed, homogeneous mixture with He, C, N, 
and O mass fractions of 0.981, $\ee{1.41}{-4}$, 0.0122, and $\ee{1.06}{-3}$ 
\citep[typical of the core composition of solar metallicity stars at the base 
of the red-giant branch that have undergone CNO-cycle burning;][]{schaller92,girardi00}.  The remaining metals are assumed to have a solar abundance pattern.  We use  OPAL radiative opacities \citep{iglesias96} calculated for this mixture for temperatures above $10^4$ K. Below $10^4$ K, we use \citet{ferguson05} opacities calculated \emph{excluding} contributions due to grains (which form beginning 
at $T \approx 2500$ K).  Where conductive opacities are relevant, we utilize 
those of \citet{potekhin99}.  As metals make negligible contributions to the 
donor's EOS \citep{chabrier97}, the pure-He EOS of \citet{winisdoerffer05} 
supplemented by radiation contributions are used.  This EOS is calculated 
using a free-energy minimization model and provides the thermodynamic quantities 
of a He fluid over a wide range of density and temperature including, in 
particular, the regimes where He undergoes thermal and pressure ionization 
state transitions.

\section{The Initial State of the Donor in White Dwarf Channel Systems}
\label{sec:initial_models}
As shown in \citet{deloye05}, evolution during the post-contact AM CVn phase 
is strongly influenced by the state of the donor at contact.  In this section, 
we determine a set of initial donor models that  encompass 
the range of donor properties at contact produced in WD channel AM CVn systems 
and discuss the expected distribution of these properties based on the 
population synthesis calculation of \citet{nelemans01a}.
    
\subsection{Determination of the Initial Donor Models}
 \label{subsec:donor_intial_structure}

To determine the range of parameters our initial donor models should cover, we utilize data from the \citet{nelemans01a} population synthesis model and the methodology of  \citet{deloye05}.  Specifically, for each WD channel system in the population synthesis model that is expected to survive initial contact \citep[see][]{nelemans01a}, we determine the proto-donor's core conditions at the beginning of the second CE using existing stellar evolution calculations performed with the EZ code \citep{paxton04} for the \citet{deloye05} study.  During the rapid CE phase, the proto-donor's central degeneracy doesn't change.  So we utilize the central degeneracy parameter---$\psi_c =  E_{F,c}/ k T_c \approx \rho_c/(\ee{1.2}{-8} T_c^{3/2}) \mathrm{K^{3/2}\, cm^3 \,g^{-1}}$, where $\rhoc$, $T_c$, $E_{F,c}$ are the central density, temperature, and electron Fermi energy---to map between the EZ stellar models pre-CE and a set of post-CE isolated He WD models calculated using with our own stellar evolution code. 

Once we determine this post-CE donor model, for each system we evolve the donor's radius and the orbital separation from this post-CE state to determine when GW emission drives the binary back into contact. The population synthesis data provides the post-CE orbital separation, $a_0$, and the initial accretor and donor masses, $\Moi$ and $\Mtwi$. We use these quantities and our single He-star tracks to calculate, self-consistently, the time elapsed in the post-CE phase, $\tcont$, before $R_2 = \RL$. This determines the donor's contact radius and $\psic$ (which we denote by $\Rtwi$ and $\psici$).  When taken over the set of population synthesis data, this procedure allows us to determine the range of $\psici$, $\Mtwi$, and $\Moi$ that occurs in the WD channel systems. 

Before discussing the distribution of donor contact parameters obtained with this procedure, we mention one complication we encountered.  In some systems, the standard CE-evolution prescription \citep[see, e.g.,][]{webbink84} predicts post-CE conditions where \emph{the donor is already in Roche contact}.  This is due to some systems having sufficiently hot proto-donors with $R_2$ significantly greater than the zero-temperature configuration. It is unclear whether such systems will cleanly exit the CE or simply be driven to a prompt merger. 

A realistic determination of this question is beyond the scope of this paper. Instead, in order to examine how this issue could influence the distribution of donor contact parameters, we considered two different, \emph{rough} criteria for determining which systems survive the CE.  In one, we removed all such systems from the population.  In the second, we looked at the evolution of $R_2$ (based on the \emph{isolated} He WD structures) and $\RL$ to determine if $R_2 < \RL$ was ever satisfied before the system was driven to $a=0$ due to GW emission.  This could happen if the proto-donor is able to contract more rapidly than GW emission decreases $\RL$.  

We found that the choice of criteria effected mainly the distribution of $\psici$ at fixed $\Mtwi$ for systems with $\Mtwi \leq 0.175 \msun$.  Taking the second criteria, About 50\% of systems with $0.10 \msun < \Mtwi \leq 0.125 \msun$ and 25\% of those with $0.125 \msun < \Mtwi \leq 0.175 \msun$ do not exit the CE by this rough criteria, where as only 4\% of those with $\Mtwi > 0.175 \msun$ are affected.  More importantly, the range of $\psici$ in the overall population is not affected by these border line systems, since  donors with $\Mtwi \gtrsim 0.2 \msun$ exhibit the full range of $\psici$ and these more massive donors all appear to exit the CE cleanly.  This general conclusion also applies to concerns about donors with $\Mtwi \approx 0.1 \msun$.  The progenitors of these donors will not have developed a distinct core/envelope structure at the start of the CE, making it unlikely that they will survive the CE-event \citep{taam00}.  Thus, excluding such systems from consideration will also not influence the $\psici$ range expected in this population.

\subsection{The Distribution of Initial Donor Parameters}
\begin{figure}
\plotone{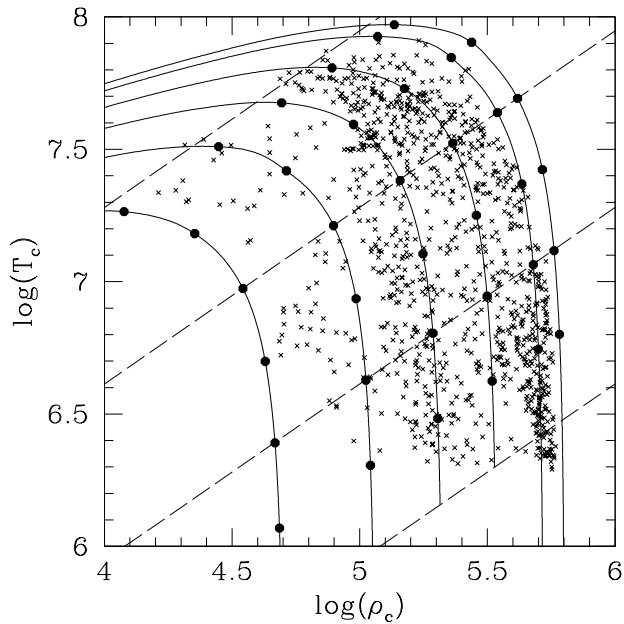}
\caption{The donors' $\rhoc$, $T_c$ at contact for each WD channel AM CVn system in the \citet{nelemans01a} population synthesis model (crosses).  For comparison, the solid lines show the $\rhoc$, $T_c$ evolution along each of our isolated He-star evolution tracks (with $M_2$ values of, from bottom to top, 0.10, 0.15, 0.20, 0.25, 0.30, and 0.325 $\msun$ ).  Lines of constant $\log \psic = $1.0, 2.0, 3.0, and 4.0 are shown by the dashed lines (top to bottom). The solid circles show the central conditions of the set of initial donor models used to begin our AM CVn phase calculations.}
\label{fig:contact_rhocTc}
\end{figure}

The results of the above exercise produces a population with $\Mtwi \approx 0.1 - 0.325 \msun$  
and $\log(\psici) \approx 1.0-4.0$ (i.e., donors are mildly to extremely degenerate).  Our determination of the donor's $\rho_c$, $T_c$ at contact for each system in the \citet{nelemans01a} population synthesis model, is shown in  Figure \ref{fig:contact_rhocTc} by crosses.  Each cross represents the starting point for a number of systems, so the density of crosses does not correspond to the number of systems starting within a given region of this parameter space. Dashed lines show lines of constant $\psic$,  while solid curves show the evolution along our isolated He WD tracks.  We have excluded systems not exiting the CE according to our second criteria above from this plot; if we were to use the first criteria, the only change here would be the loss of  most systems with $0.1 \lesssim \Mtwi \lesssim 0.15 \msun$ and $1.0 \lesssim \log(\psici) \lesssim 1.5$.  The solid circles in this figure show the central conditions of the initial donor models we use in subsequent calculations.

\begin{figure}
\plotone{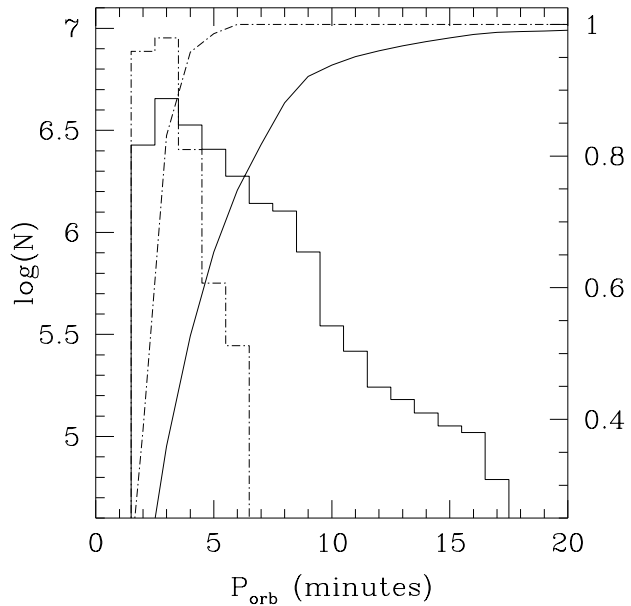}
\caption{The contact $\Porb$-distribution as determined by self-consistent treatment of the donor's cooling before contact (solid histogram) compared to assuming all donors are fully degenerate at contact (dash dotted histogram).  The number of systems in each bin gives the total number of current WD channel AM CVn systems in the \citet{nelemans01a} population synthesis model whose contact $\Porb$ fell within each bin.  The solid and dashed curves show the corresponding cumulative fractional distribution of contact $\Porb$.}
\label{fig:contact_porb}
\end{figure}

We show the distribution in orbital period, $\Porb$, at contact (integrated over all $\Mtwi$) in Figure \ref{fig:contact_porb} with the solid line. In this plot, the number, $N$,  in each bin represent the total number of WD channel AM CVn systems in the \citet{nelemans01a} population synthesis model that make contact within each $\Porb$ range.  For comparison we also show with the dashed line the distribution that results from assuming fully degenerate donors. We take $\Porb$ at contact to equal $53.5 (\Rtwi/ 0.1 \rsun)^{3/2} (\Mtwi 0.1 \msun)^{-1/2}$ minutes and calculate the fully degenerate $\Rtwi$ using a fit to the \citet{deloye03} zero-temperature He mass-radius relation:
\begin{multline}
R_{2,\mathrm{DB}} = 0.005593 - \frac{\ee{1.7616}{-5}}{M_2}\\ + 0.004643\,M_2 - 0.008298 \ln M_2 \,, 
\end{multline} 
which is accurate to better than 3\% over the range $\ee{9}{-4} \msun \leq M_2 \leq 0.45 \msun$.  The fully degenerate distribution peaks at $\Porb = 3.0$ minutes with a cut-off at 6 minutes.  On the other hand, roughly 25\% of systems within our current calculation make contact at a $\Porb > 6$ minutes and 5\% at $\Porb > 11$ minutes. Additionally, there is a small tail of systems making contact out to $\Porb = 35$ minutes (not shown in Figure  \ref{fig:contact_porb}). If we remove all systems with post-CE $R_L < R_2$, this tail extends out to 22 minutes, but otherwise the contact $\Porb$-distribution is almost unchanged.

\begin{figure}
\plotone{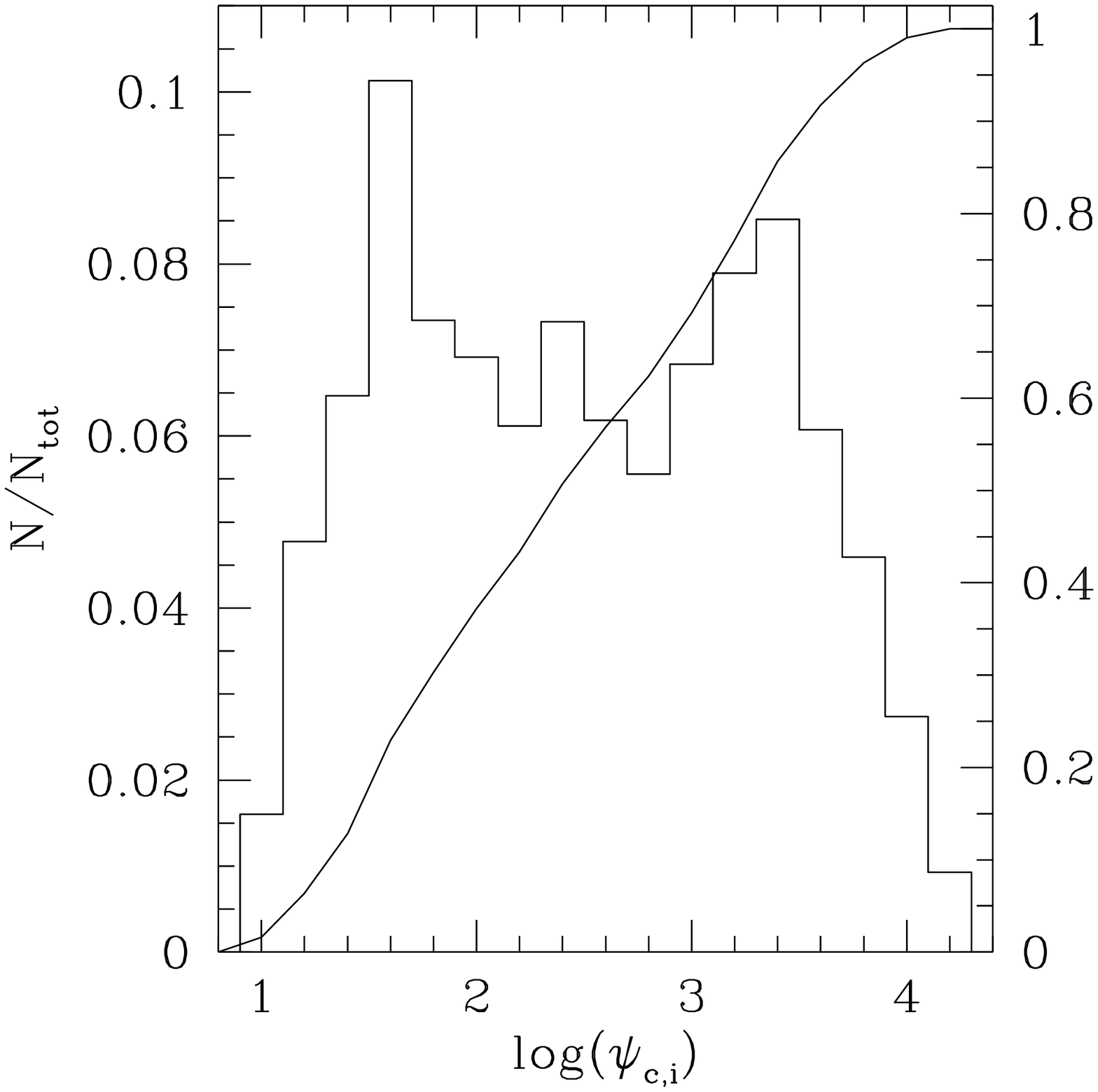}
\caption{The $\psici$ distribution, integrated over all $\Mtwi$, normalized to the 
total number of systems in the population model.  The solid curve provides the 
cumulative distribution. Unlike the contact $\Porb$ distribution in Figure 
\ref{fig:contact_porb}, which is strongly peaked at shorter periods, the $\psici$ 
distribution is relatively flat between $\log(\psici)\approx 1.4$ and 3.6.  
}
\label{fig:psici_hist}
\end{figure}

The corresponding $\psici$-distribution is shown by the histogram in Figure \ref{fig:psici_hist}.  The solid curve in this figure gives the cumulative distribution of $\psici$ values.  Noteworthy is the fact that, unlike the contact $\Porb$-distribution, which is strongly biased towards short $\Porb$, the $\psici$-distribution is approximately flat between $\log(\psici)\approx 1.4$ and 3.6.  The difference in the two distributions is due to the dependence of $R_2$ on $\psici$: $R_2$ varies rapidly with $\psici$ for $\psici \sim 1-10$, but only more slowly for $\psici \gtrsim 100$. Thus there is not as strong an a priori theoretical preference for large $\psici$ values in the WD channel population as might be expected from the contact $\Porb$-distribution. The $\psici$-distribution is essentially unchanged when removing all systems with post-CE $R_L < R_2$ from consideration.

\section{The Donor's Structural and Thermal Evolution}
 \label{sec:ev_general}
In order to adequately sample the range of initial donor properties derived in \S \ref{sec:initial_models} in our subsequent calculations, we considered AM CVn evolution starting from 36 different initial donor models (solid circles in Fig. \ref{fig:contact_rhocTc}) with $\Mtwi$ values of 0.10, 0.15, 0.20, 0.25, 0.30, and 0.325 $\msun$ and $\log(\psici)$ values of 1.1, 1.5, 2.0, 2.5, 3.0, and 3.5.  The coupled binary and donor evolution were calculated for two $\Moi$ for each donor model.  These $\Moi$ were chosen so that $\Mtot = \Mtwi + \Moi$ equalled one of three values---0.5, 0.825, or 1.325 $\msun$.

To present our results, we first discuss the donor's evolution in detail for several representative cases.  We then examine the entire range of our calculations, detailing how and in which evolutionary phases different initial conditions affect the system's parameters.  One of our main results is that the donor's outer boundary condition is dominated by irradiation from the accretion flow. We also, therefore, present additional calculations including the effects of external irradiation and discuss the impact of this additional physics on the system's evolution.

\subsection{The  AM CVn Donors' Evolutionary Phases}
\label{sec:ev_rep}
\begin{figure}
\plotone{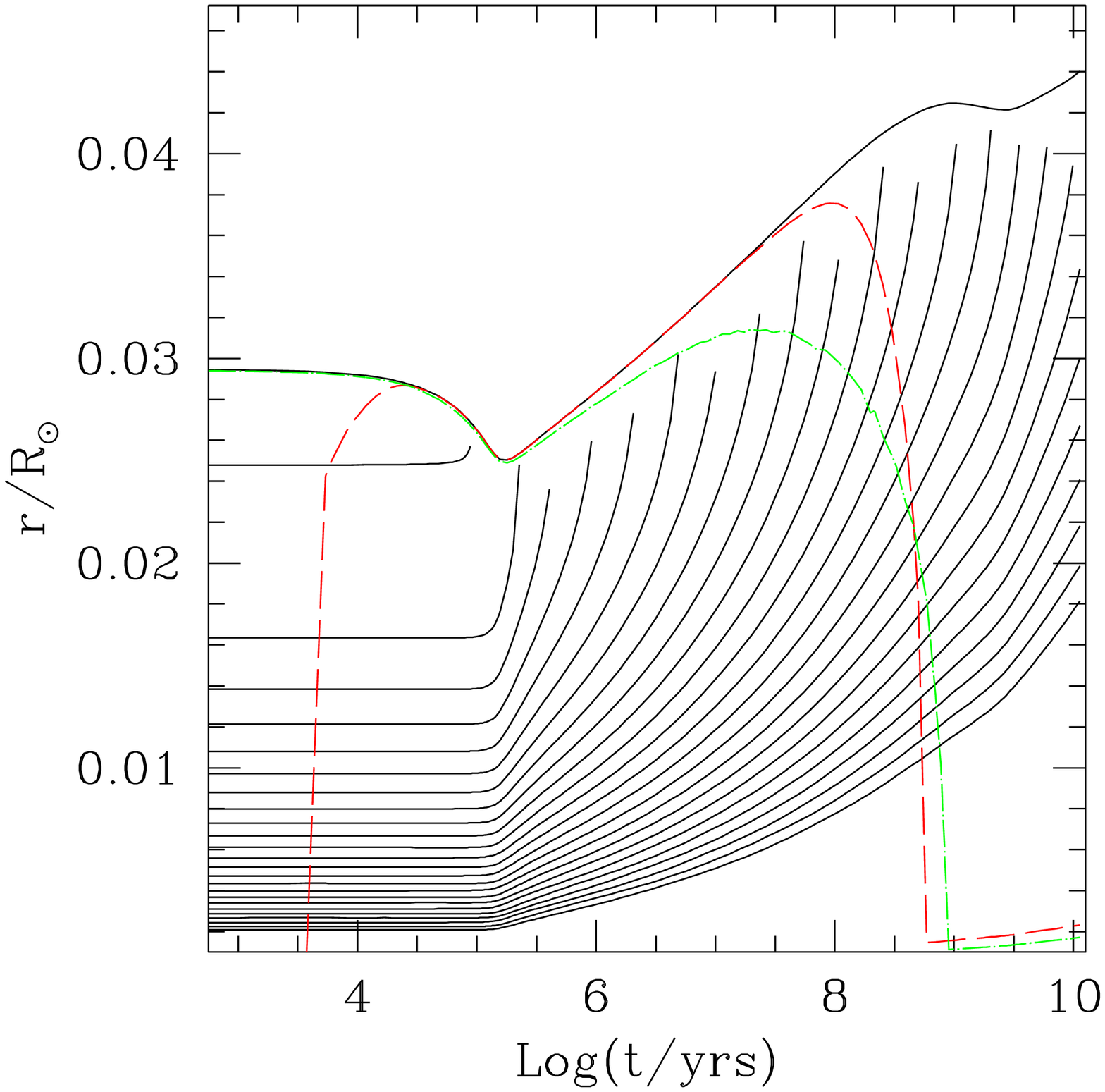}
\caption{The time evolution of the radius at constant $m$ (black lines, the bottom line shows $m=0.001 \msun$ with higher lines spaced by $\Delta \log m = 0.1$; the top-most black lines traces $R_2$), the $r$ at which $\tauth=\taum$ (dashed red line, below this line $\tauth > \taum$), and the lower boundary of the outer convective zone (dashed green line). The initial conditions here are $\Mtwi=0.2\,\msun$, $\log(\psici)= 2.0$, $\Rtwi=0.0294\,\rsun$, and $\Moi=0.3\,\msun$. }
\label{fig:R_lagrange_ev}
\end{figure}

\begin{figure}
\plotone{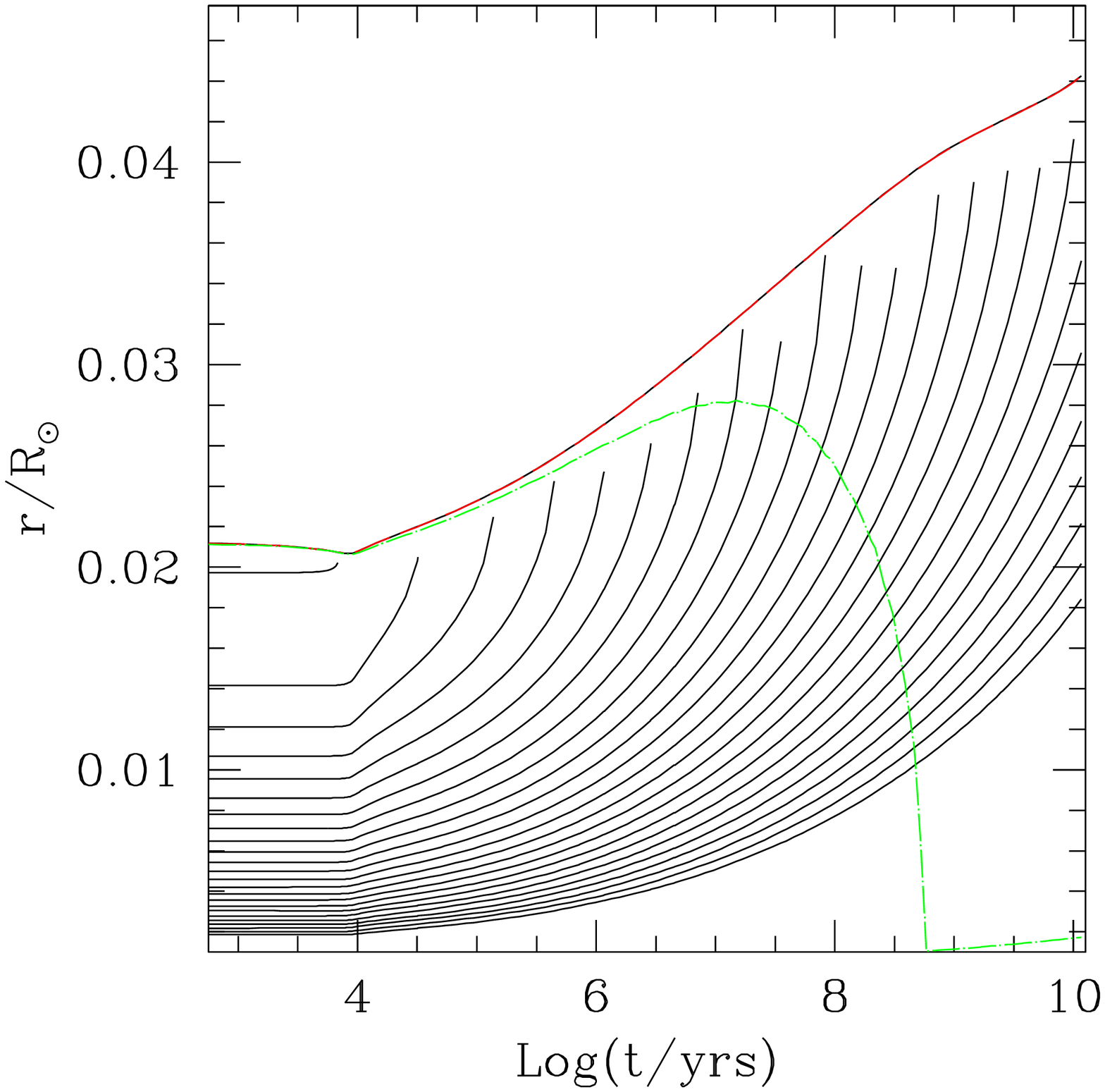}
\caption{Same as Figure \ref{fig:R_lagrange_ev}, but with a more degenerate donor; 
the initial conditions in this system are $\Mtwi = 0.2\,\msun$, $\log(\psici) 
= 3.5$, $\Rtwi = 0.0212\,\rsun$, and $\Moi = 0.3\,\msun$}
 \label{fig:R_lagrange_evb}
\end{figure}

Figure \ref{fig:R_lagrange_ev} shows a summary of the donor's evolution in a representative calculation.  The initial conditions there were $\Mtwi=0.2\,\msun$, $\log(\psici)=2.0$, $\Rtwi=0.0294\,\rsun$, and $\Moi=0.3\,\msun$. This model illustrates the range of donors' responses to mass loss; for comparison, Figure \ref{fig:R_lagrange_evb} shows a similar summary for a more degenerate, $\log(\psici)=3.5$,  donor.  In Figure \ref{fig:R_lagrange_ev}, the solid black lines (except for the top-most) show the $r$ evolution at fixed $m$ as a functions of time (the calculations are started with $t=0$ at an $a$ such that $\RL \approx R_2+30 H_P$). The top-most solid line traces out the $R_2$ evolution.

The other lines in this plot indicate various physical conditions in the donor.  All donor models have, to some extent, an outer convective region; the dashed-dotted green line indicates the lower radius to which this convective zone penetrates. The dashed red line indicates the $r$ at which the local thermal time, $\tauth$, equals the local mass loss time, $\taum$, where these two quantities are defined as
\begin{equation}
\tauth = \frac{\int_0^{m'} c_P T dm''}{L} \,,
\end{equation}
\begin{equation}
\taum = \frac{m'}{\Mdot}\,,
\end{equation}
and $m' = M-m$ is the mass exterior to the specified location in the donor.  Regions below the dashed red line in Figure \ref{fig:R_lagrange_ev} satisfy $\tauth > \taum$. The evolution in layers where $\tauth \gg \taum$ is dominated by the nearly adiabatic advection to lower $P$, while layers where $\tauth  \ll \taum$ are able to adjust their thermal structure in response to mass loss almost instantaneously.

The results shown in Figure \ref{fig:R_lagrange_ev} are typical of our calculations in that the evolution there can be divided into three phases.  In the first phase, during which $\Mdot$ grows from zero to its maximum, $R_2$ decreases towards a minimum value in response to mass loss.  In Fig. \ref{fig:R_lagrange_ev}, this phase lasts until $t \approx \ee{2}{5}$ yrs.  The second phase begins as the $R_2$ evolution reverses and the donor begins expanding in response to mass loss. This expansion phase corresponds to what is normally considered the AM CVn phase of evolution. By this point, $\tauth \gg \taum$ throughout the donor so that the donor responds adiabatically to the mass loss. Note that the donor enters this adiabatic phase well before the $R_2$ evolution reverses (see below).  At some point during the expansion phase, the $\tauth=\taum$ line begins moving inward again.  This eventually leads to the start of the third phase, where the donor is able to cool and contract. This is seen at $t \approx 10^9$ yrs in Fig. \ref{fig:R_lagrange_ev}. This contraction ends once the donor has shed sufficient  entropy to reach its fully degenerate configuration, even as it continues to cool (evolution beyond $t \approx \ee{2-3}{9}$ yrs in Fig. \ref{fig:R_lagrange_ev}).

The first and third phases of the donor's evolution are qualitatively new results resulting from our more complete treatment of the donor's physics.  A more detailed discussion of these phases follows. 

\subsubsection{The $\mdot$ Turn-on Phase: Evolution to $\mdot$-Maximum}
 \label{sec:donorev_preR2min}
Once the WD channel AM CVn donors make contact, our calculations show they all 
begin a phase of radius contraction, a result not anticipated by prior modelling 
\citep[e.g.,][]{nelemans01a,deloye05}.  Here, we discuss how the donor's structure determines the duration and extent of this initial $R_2$ contraction phase.

The basic picture of the donor's response to mass loss can be seen from Figs. 
\ref{fig:R_lagrange_ev} and \ref{fig:R_lagrange_evb}.  As $M_2$ decreases, 
mass elements move to lower $P$ and $\rho$, producing the expansion in $r$ at 
fixed $m$ seen in the evolution of the interior black lines. In Appendix 
\ref{app:R_response} we show that this expansion is most significant near the 
surface and that the only contribution to $R_2$ contraction comes from the 
surface term in equation (\ref{eq:drdmint}). In other words, mass lost from the donor takes with it its contribution to $R_2$, tending to produce contraction.  The overall $R_2$ evolution then depends on whether the underlying layers expand sufficiently to compensate for this lost radius contribution.

Equation (\ref{eq:drdm_Pint}) shows that this underlying expansion depends on the two quantities: $\chi_T \equiv (\partial \ln P/\partial \ln T)_{\rho}$ and $\del - \del'$, where $\del' = (\partial \ln T/\partial \ln P)_m$ describes the actual thermal evolution of a mass element as it is advected to lower $P$. In layers where either of these quantities tends towards zero, the $\rho (P)$ profile remains constant under advection, producing no net change in the relative contribution to $R_2$ from that region. This can occur for strongly degenerate plasmas, where  $\chi_T \approx 0$, or when the advected mass element arrives at a lower $P$ with the same entropy as the material it replaces.  Only when the advected mass elements arrive with lower entropy ($\del' > \del$) is the relative $R_2$ contribution reduced.  If the latter case dominates in the outer layers, this can lead to a net decrease in $R_2$.

Degeneracy effects never dominate in our donors outer layers---at least during this contact phase---so the $R_2$ evolution depends only on the mode of heat transport in the outer layers and on the ordering of $\tauth$ and $\taum$.  When $\tauth \ll \taum$, heat transport has sufficient time to redistribute entropy so that $\del' \approx \del$, and little $R_2$ evolution occurs.  Once $\taum \lesssim \tauth$, so that $\del' \rightarrow \delad$,  the $R_2$ evolution begins to depend on the background entropy gradient.  In convective regions, $\del$ is essentially equal to $\delad$, so that adiabatic advection leads to minimal local contributions to $R_2$ evolution.  In radiative regions, however, $\del < \delad$; \emph{thus only for radiative surface regions during roughly adiabatic advection will a net $R_2$-contraction occur} \citep[see, e.g.,] [for another, qualitative, discussion of this]{faulkner76}. The rate of the $R_2$ contraction increases with the entropy of the layer and with the steepness of the background entropy profile.

The structure of our donor's outer layers consists of a superficial radiative layer overlying a thin outer-convective zone, followed by another radiative region that extends to the stellar centre.  The thickness of outer-convective region increases with donor degeneracy.  The entropy profile of the inner radiative region is very steep near its outer boundary; moving inward, this entropy profile tends to flatten out rather abruptly.  With this general interior structure in mind, the $R_2$ evolution trends seen in Figs. \ref{fig:R_lagrange_ev} and \ref{fig:R_lagrange_evb} can be understood from the above discussion.

For the $\log(\psici)=2.0$ donor (Fig. \ref{fig:R_lagrange_ev}) , as long as $\tauth \ll \taum$, the $R_2$ evolution is minimal.  Once this inequality is reversed, the $R_2$ evolution starts to accelerate and a rapid contraction ensues due to the steep entropy profile in the donor's radiative photosphere.  By this time, the underlying layers are advected nearly adiabatically, and the entropy gradient near the surface is continuously decreasing. This leads to the decreasing rate of $R_2$ contraction. Finally the entropy profile becomes sufficiently shallow that the expansion of the underlying layers takes over, and $R_2$ begins increasing.

The evolution in Fig. \ref{fig:R_lagrange_evb} is similar, but here the evolution is adiabatic ($\tauth \gg \taum$)throughout.  The $R_2$ contraction rate is slower both because the outer convective zone is thicker and the underlying radiative region has both a lower entropy and a shallower entropy profile. This also results in less overall $R_2$ contraction. Additionally, $R_2$ decreases significantly only once a sufficient amount of mass, $\delta m$, has been lost for the pressure perturbation in the outer layers, $\delta P \approx G \Mtwi \delta m/(4 \pi \Rtwi^4) \approx \ee{9}{20}\, \mathrm{dyne\,cm^{-2}}\, (\delta m/\Mtwi)\,(\Mtwi/0.1\msun)^2 (0.01 \rsun/\Rtwi)^4$, to be of order the initial $P$ at the top of the radiative region. 

\begin{figure}
\plotone{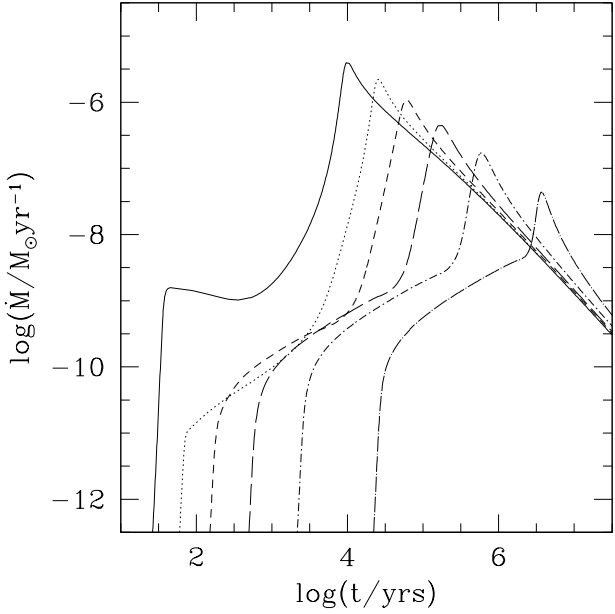}
\caption{The $\mdot (t)$ evolution for AM CVn systems with $\Mtwi=0.2$ and 
$\Moi=0.3 \msun$. Systems with donors having $\log(\psici)=$ 3.5, 3.0, 2.5, 
2.0, 1.5, and 1.1 are shown by the solid, dotted, short-dashed, dashed, short-dash 
dotted, and dash dotted lines, respectively.  The corresponding $\Rtwi$ = 0.0212, 
0.0225, 0.0248, 0.0294, 0.0380, 0.0561 $\rsun$, respectively }
\label{fig:turnon_mdot_t}
\end{figure}

The relative rate of $R_2$ and $\RL$ evolution---characterized by $\nrtw \equiv (d \ln R_2/d \ln M_2) $ and $\nrl \equiv  (d \ln \RL/d \ln M_2)$---determines that of $\mdot$.  In all our evolution models, $\nrl \gg \nrtw$ initially. However, since 
\begin{equation}
\nrl  = 2 \left[ \left(\JdotJ\right)_{\mathrm{GW}} \frac{M_2}{\mdot_2} + q - \frac{5}{6} \right]\,,
\end{equation} 
$\nrl$ rapidly decreases with growing $\mdot$.  Once $\nrl \approx \nrtw$, these two quantities tend to track each other closely due to the sensitive $\DelR$ dependence of $\mdot$. Thus, when  the $\nrtw$ evolution is  smooth, the $\mdot$ evolution is also.  

This can be seen in Figure \ref{fig:turnon_mdot_t}, which shows the $\mdot$ time evolution for a set of models with $\Mtwi=0.2 $, $\Moi=0.3 \msun$ and differing $\psici$ (indicated by line style with $\psici$ decreasing left-to-right).  The four lowest $\psici$ donors have very thin outer convective zones  and their corresponding $\nrtw$ evolution is smooth.  The two highest $\psici$ donors (dotted and solid curves) have their $R_2$ response dominated, at first, by their thicker outer convective zones.  Once $\delta m$ begins probing the underlying radiative region, their $\nrtw$ suffer rapid changes in slope.  This transition from convective to radiative dominated $R_2$ response produces the non-smooth and non-monotonic behaviour seen in the dotted and solid curves.

Figure \ref{fig:turnon_mdot_t} illustrates other general trends.  Initially, $\RL > R_2$ in all cases.  While this remains true, the $\mdot$ growth is approximately exponential (corresponding to the initial, steep increase in $\mdot$).  Lower $\psici$ donors produce slower $\mdot$ growth both due to their larger contact $a$ (reducing the system's $\dot{J}/J$) and larger $H_P/R_2$; the latter tends to be the more significant factor. Since the $\mdot$ required to produce $\dot{a} > 0$ is greater than $\mdot_0$ in all the systems we considered, all experience a phase where $\RL < R_2$.  Once this occurs, the $\mdot$ growth slows as $\mdot$ now depends on the donor's non-exponential, sub-photospheric $\rho$-profile (see equation \ref{eq:m2dot_thick}).  This results in the turn-over in $\mdot$ growth seen at $\mdot \sim 10^{-10}$-$10^{-9} \msun$ yr$^{-1}$ . The final upturn in $\mdot$ before each maximum results from a rapid increase in $\mdot_0$ via the increasing $\rho$ of the surface layers.  This results from the adiabatic advection of lower entropy material to the surface and corresponds to the final phase of rapid $R_2$ contraction discussed above.

The extrema in $R_2$, $a$, and $\Mdot$, which mark the end of the turn-on phase, do not occur simultaneously.  The $R_2$-minimum happens first and the subsequent expansion contributes to accelerating the $\mdot$ growth. The $a$-minimum occurs once $\Mdot$ reaches a critical value, $\Mdot_{\mathrm{crit}} = - (\dot{J}/J)_{\mathrm{GW}} M_2/(1-q)$ to produce $\dot{a}=0$.  The larger $\Jdot/J_{\mathrm{GW}}$ in systems with more degenerate donors produces higher $\Mdot_{\mathrm{crit}}$.  Eventually, the $a$-expansion leads to a decreasing $\DelR$, and $\mdot$ reaches its maximum. By $\mdot$-maximum, the donor has lost between $\delta m \approx (0.02-0.2) \Mtwi$. Less degenerate donors suffer the greater mass loss.  Overall the turn-on phase lasts between $\sim 10^4$ and $\sim 10^6$ yrs. 

\subsubsection{The Donor's Cooling Phase}
\label{sec:donorev_postR2min}
By the time the system has evolved past $\mdot$-maximum, the donor is evolving adiabatically in response to mass loss.  Prior work on the evolution of AM CVn systems in this phase has neglected the donor's thermal evolution and has either relied on predetermined $M$-$R$ tracks \citep{nelemans01a, farmer03} or has assumed the donors continue to evolve adiabatically indefinitely \citep{deloye05}. Our current calculations show that at late-times, the assumption of adiabatic evolution becomes invalid (as can be seen by the evolution of $\taum=\tauth$  in Fig. \ref{fig:R_lagrange_ev}).

Why the donor's adiabatic evolution ends can be understood by considering how $\taum$ and $\tauth$ evaluated at $m' = M_2$ evolve under mass loss.  The quantity 
\begin{equation}
\der{\ln \taum}{\ln M_2}\Big\arrowvert_{m'=M_2} = 1 - \der{\ln \Mdot}{\ln M_2}\,.
\end{equation}
For GW driven, conservative mass transfer, the latter derivative can be written as
\begin{equation}
\der{\ln \Mdot}{\ln M_2} = 2 - q - 4 \der{\ln a}{\ln M_2} \approx 2 - 4\left(\nrtw - \frac{1}{3} \right)\,,
\end{equation}  
where the final approximation is good when $M_2 \ll M_1$.  In the same regime, 
\begin{equation}
\der{\ln \taum}{\ln M_2}\Big\arrowvert_{m'=M_2} \approx  4\left(\nrtw - \frac{1}{3} \right) -1 \,.
\end{equation}

It is useful to rewrite $\tauth = \int_0^{M_2} c_P T dm'/L$ as  $c_P' T_c M_2/L$ where $c_P' =  \int_0^{M_2} c_P T dm'/(T_c M_2)$ since both $c_P'$ and $L$ vary only by a factor of a few during the adiabatic phase.  In contrast, $M_2$ and $T_c$ vary by roughly one and two orders of magnitude, respectively.  Holding $c_p'$, $L$ constant, $d \ln \tauth/d \ln M_2 \approx 1 + d \ln T_c/d \ln M_2$, and we can use hydrostatic balance to write
 \begin{equation}
\der{\ln T_c}{\ln M_2}\Big\arrowvert_{m'=M_2} = \left(\der{ln T_c}{\ln P_c}\right) \left(\der{\ln P_c}{\ln M_2}\right) = \nabla_{\mathrm{ad},c} (2 - 4 \nrtw)\,, 
\end{equation}
where $\nabla_{\mathrm{ad},c}$ characterizes the $T_c$ evolution during the adiabatic phase.  

\begin{figure}
\plotone{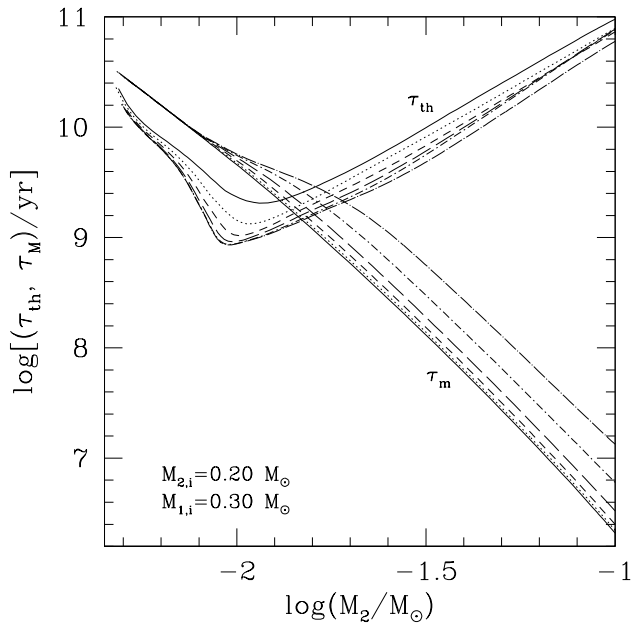}
\caption{The evolution of $\tauth$ and $\taum$ (at $m'=M_2$) as a function of $M_2$ for the set of AM CVn models with $\Mtwi = 0.2 \msun$ and $\Moi = 0.3 \msun$ shown in Fig \ref{fig:turnon_mdot_t}.  The various lines styles have the same meaning as in that figure.}
\label{fig:tauthm_m}  
\end{figure}

Thus once $M_2 \ll M_1$ in the adiabatic phase, $\tauth/\taum 
\propto M_2^{\alpha}$ where
\begin{equation}
\alpha = 4 \nrtw - \frac{10}{3} + \nabla_{\mathrm{ad},c} (4 \nrtw -2)\,.
\end{equation} 
Typically, $\nrtw \approx -0.3$-$ -0.2$ and $\nabla_{\mathrm{ad},c} \approx 
0.36$-$ 0.39$, so that $\alpha \approx -5.7$-$ -5.2$ and leading a rapid evolution 
of $\tauth/\taum$ as $M_2$ is reduced.  In the early portion of the 
adiabatic phase, where $M_2 \sim 0.1 \msun$, $\tauth/\taum \sim 10^3-10^5$ 
and this scaling then gives $\tauth/\taum \approx 1$ when $M_2 \approx 
0.01-0.03 \msun$.  

We compare this prediction to our numerical calculations in Fig. \ref{fig:tauthm_m}, 
where we show the computed evolution of both time-scales for our models with 
$\Mtwi = 0.2 \msun$, $\Moi = 0.3 \msun$.  The ratio $\tauth/\taum \approx 1$ 
at $M_2=0.01-0.02$ in all cases, in excellent agreement with the scaling 
relation.  Beyond this point, the donor is able to shed entropy, 
allowing less degenerate donors  to contract towards a fully degenerate configuration.

\subsection{Impact of Initial Conditions on Donor Evolution}
 \label{sec:ev_noirr}
\begin{figure}
\plotone{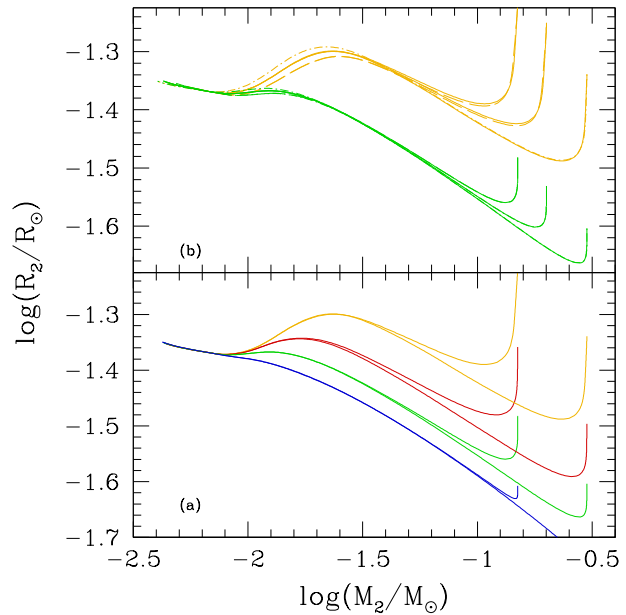}
\caption{The dependence of the $R_2$-evolution on initial binary parameters.  In both panels, colours indicate the donor's initial $\log(\psi_{c,i})$: 1.1 (yellow), 1.5 (red), 
2.0 (green), and 3.0 (blue). Line style indicates the binary's $\mtot$: 0.500 
(dashed), 0.825 (solid), and 1.325 $\msun$ (dot short-dashed). Panel (a) compares 
the $R_2$-$M_2$ evolution for systems with $\mtot=0.825 \msun$ to examine the 
impact of initial degeneracy and donor mass. We show evolution for systems with 
$\Mtwi=$ 0.15, and 0.3 $\msun$ and the initial degeneracy listed above.  
Panel (b) examines the impact of varying $\Mtot$ on the $R_2$ evolution.  We 
show comparisons at two $\psi_{c,i}$ to illustrate how the magnitude of $\Mtot$ 
effects are reduced at higher donor degeneracy. }
\label{fig:RM_binevcomp}
\end{figure}

We now explore how the donor and binary evolution varies with initial conditions. 
In Fig. \ref{fig:RM_binevcomp}, we show the $R_2(M_2)$ evolution for several 
sets of calculations. In this figure, different colours correspond to different 
initial donor degeneracy, while different line styles indicate different $\Mtot$. 
The evolution along each track is from right to left, with the donors first 
evolving steeply downward in $R_2$. The amount $R_2$ decreases prior to the 
$R_2$-minimum depends on $\psici$.  Less degenerate donors contract to a greater 
extent during the turn-on phase as discussed in \S \ref{sec:donorev_preR2min}. 

In panel (a) of Fig. \ref{fig:RM_binevcomp} the $R_2$ evolution for a set of systems with $\Mtot=0.825 \msun$ are displayed.  During the turn-on phase and early expansion phase, $R_2(M_2)$ depends on both $\psici$ and $\Mtwi$. At fixed $\psici$ different $\Mtwi$ have differing $s(m)$ profiles, and this is reflected in the $R_2$ evolution. The entropy differences are more significant at larger $m$, producing the tendency towards convergence as $M_2$ is reduced. Near the local $R_2$-maximum (at $M_2 \approx 0.01-0.03$), track convergence is furthered by donor cooling, which erases any remaining initial $s$-profile information.  By this point, the donors cool and contract along tracks parametrized by $\psici$. Cooling continues and $R_2$ contraction slows as the donors become increasingly degenerate and finally reach the fully degenerate \MR relation.  

In panel (b) of Fig. \ref{fig:RM_binevcomp}, we consider the dependence of the 
$R_2$ evolution on $\Mtot$. During the turn-on phase, the donor evolution is 
essentially independent of $\Mtot$.  Only for the very lowest degeneracy 
donors is this not the case. These donors are able to cool somewhat during the turn-on
phase and systems with lower $\Mtot$ produce slower $\Mdot$ growth, allowing 
greater cooling before mass loss becomes adiabatic. The larger $\Mdot$ in higher 
$\Mtot$ systems also extends the duration of the adiabatic expansion phase to 
lower $M_2$ and larger $R_2$.  The $R_2$ path a system follows during its 
later contraction phase is thus parametrized by $\psici$ and $\Mtot$. 

\begin{figure}
\plotone{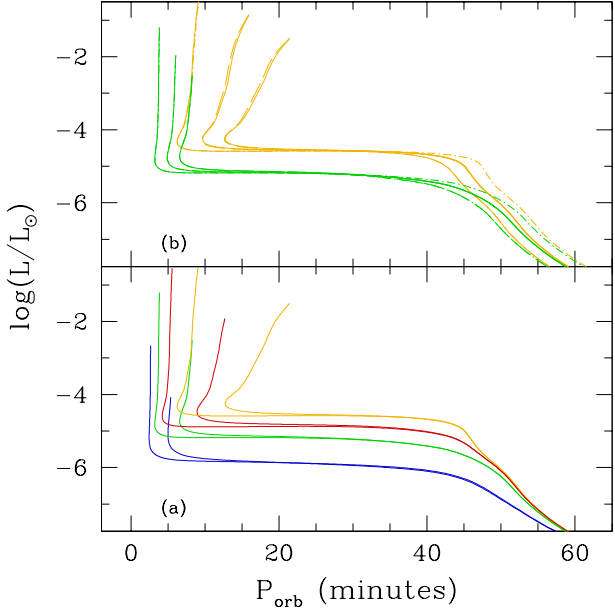}
\caption{The dependence of the donor's $L$-evolution on initial binary parameters.  The line styles and colours have the same meanings as in Fig. \ref{fig:RM_binevcomp}.  As in that figure, panel (a) shows the effects of varying $\Mtwi$ and $\psici$ at fixed $\Mtot$, while panel (b) examines the effects of varying $\Mtot$, $\Mtwi$ at fixed $\psici$. Here $\Mtwi$ has values of 0.15, and 0.3 $\msun$.  Evolution proceeds downward along each track and $\Mtwi$ can be inferred from the initial $\Porb$ within each set of constant $\psici$ tracks, with $\Mtwi$ decreasing with increasing $\Porbi$.}
\label{fig:TeffLPo_binevcomp}
\end{figure}

In Fig. \ref{fig:TeffLPo_binevcomp} the evolution of the donor's $L$ is shown as a function of $\Porb$, with evolution proceeding downward along each track.  The $L$-evolution reflects the donor's evolutionary stages.  The significant decrease during the turn-on phase (while $\Porb$ is decreasing) is due primarily to the removal of mass from the steep $s$-profile region (see Appendix \ref{app:L_response}). Once the steep $s$-gradient material has been removed, the $L$ evolution slows along with the $R_2$ contraction.  In the adiabatic expansion phase, $L$ roughly plateaus at a level set by the core entropy profile. That is, $L$ during this phase is parametrized by $\psici$ and $\Porb$. Variations in the $s$ profiles due to $\Mtwi$ differences are reflected in slight differences between tracks sharing $\psici$. During this phase, $L \approx $10$^{-6}$--10$^{-4}$ $\lsun$, while $\Teff$ also plateaus at values $\approx $1000--1800 K. The slow evolution in $L$ ends once donor cooling sets in, producing the rapid decline starting between $\Porb \approx $40-50 minutes.  Panel (b) shows that a larger $\Mtot$ extends the $\Porb$ at which donor cooling start (via an increased $\Mdot$); $\Mtot$ thus parametrizes the $L$ evolution during the cooling phase and beyond, with higher $\Mtot$ producing higher $L$ at fixed $\Porb$.

\begin{figure}
\plotone{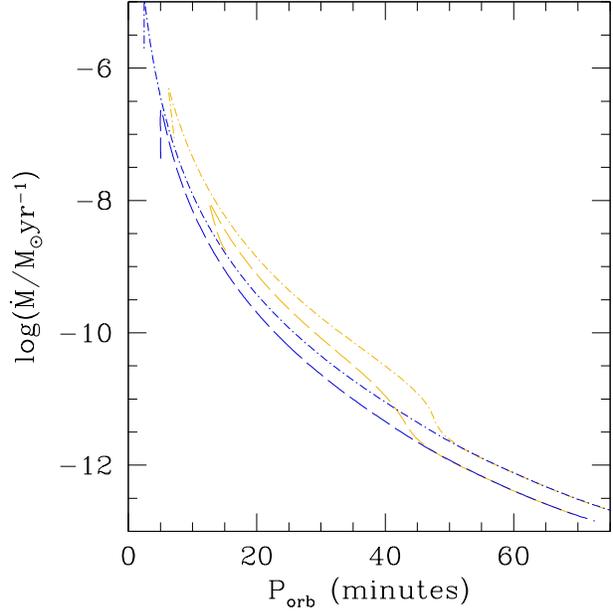}
\caption{The dependence of the $\Mdot$-$\Porb$ relation on initial binary parameters. The line styles and colours have the same meanings as in Fig. \ref{fig:RM_binevcomp}. For clarity, we display the evolution of two ($\Moi$, $\Mtwi$) pairs: (0.350, 0.150), and (1.025, 0.300), where values are in $\msun$ and values of $\log(\psici)$ of 3.0 and 1.1. The figure focuses on the $\Mdot$ evolution post $\Mdot$-maximum. Evolution is first upward along the vertical sections of each track, then outward in $\Porb$ as $\Mdot$ decreases. The range of $\Mdot$ at fixed $\Porb$ displayed here represents the range available to WD channel AM CVn binaries \emph{given our assumptions in determining the donor's initial entropy}.  %
}
\label{fig:MdPo_binevcomp}
\end{figure}
The differences in $R_2(M_2)$ evolution with $\Mtot$ and $\psici$ are reflected 
in the binary's $\Mdot(\Porb)$ evolution as shown in Fig. \ref{fig:MdPo_binevcomp}. 
For clarity, the evolution is shown for only four systems; these systems do, however, 
show the range of phase-space covered by the overall population in this study. 
We show models with $\log(\psici) =$3.0 and 1.1, the latter providing a reasonable 
lower limit to the degeneracy expected in this population. 

Although not shown by this plot, the $\Mdot$-$\Porb$ evolution during 
turn-on is determined primarily by $\Mtwi$ and $\psici$; $\Mtot$ influences only 
the $\Mdot$ maximum at large $\psici$.  Once the systems 
have evolved into the expansion phase, the $\Porb$-$\Mdot$ evolution depends 
only on $\Mtot$ and $\psici$, since donors with the same $\psici$ eventually follow the 
same $R_2(M_2)$ evolution.  For GW-driven $\Jdot$, at fixed 
$\Porb$, $\Mdot \propto M_1^{2/3} M_2^2$ (for $M_2 \ll M_2$) so that the variations 
in the donors' $R_2$ evolution produce a larger change in $\Mdot$ than do $\Mtot$ 
variations.  This trend is shown in Fig. \ref{fig:MdPo_binevcomp}, where the 
family of curves with $\log(\psici)=1.1$ lie significantly above those with 
$\log(\psici)=3.0$. Within each set of curves at fixed $\psici$, the spread 
in $\Mdot$ is caused by $\Mtot$ differences.  Once donor cooling sets in, all tracks 
collapse toward the fully degenerate ones. 
Systems with lower $\psici$ and larger $\Mtot$ reach the fully degenerate tracks 
at longer $\Porb$. After this point, the $\Mdot$-$\Porb$ relation is parametrized 
exclusively by $\Mtot$. 

\subsection{Evolution with Irradiative Feedback \label{sec:ev_irr}}
\subsubsection{Comparison of the Irradiative and Donor's Intrinsic Fluxes}
The donor's thermal evolution does not occur in isolation, but in the radiation 
bath provided by the flux from the accretor and disk.  As the compressional heating 
luminosity generated in the accretor is always much less than the accretion 
luminosity \citep{bildsten06}, the external flux seen by the donor is dominated 
by the accretion light.  To determine how important this external flux may be to the donor's evolution, we calculate the accretion luminosity, $\Lacc$, using
\begin{equation}
\Lacc = \Mdot (\phi_{L1} - \phi_{R1})\,,
\label{eq:Laccdef}
\end{equation}
where $\phi_{L1}$, $\phi_{R1}$ are the gravitational potential at the inner 
Lagrange point and the accretor, respectively \citep{han99}, and  then compare  $\Tirr = (\Lacc/4 \pi \sigma a^2)^{1/4}$ to the donor's $\Teff$.   We use equations (14)-(16) 
of \citet{han99} to calculate $\phi_{L1}$, $\phi_{R1}$ (noting these authors interchange our definitions of $M_1$ and $M_2$).
\begin{figure}
\plotone{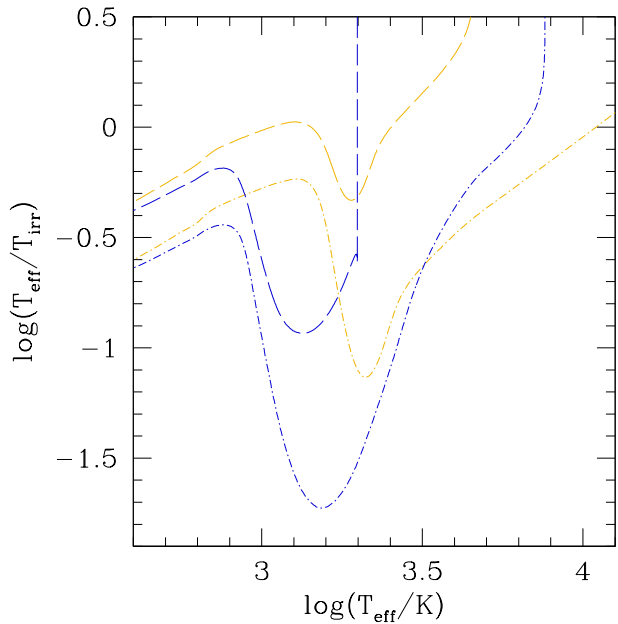}
\caption{A comparison between the $\Teff$ of non-irradiated donors and the $\Tirr$ 
produced by the system's accretion flow. We show the same set of models as in 
Figure \ref{fig:MdPo_binevcomp}.  Line styles and colours have the same meaning 
here as in that Figure.  For the entire range of donors, irradiation dominates 
the donor's outer boundary condition by the time $\Mdot$ has grown to its maximum 
value.}
\label{fig:Tirrcomp}
\end{figure}

The evolution of $\Teff/\Tirr$ versus $\Teff$ for the evolution tracks from 
Figure \ref{fig:MdPo_binevcomp} is shown in Figure \ref{fig:Tirrcomp}. During 
the turn-on phase, $\Tirr$ increases rapidly, and by $\Mdot$-maximum, $\Tirr$ has grown larger than $\Teff$ by a up to a factor of 50.  In all but the lowest $\Mtot$ and 
$\psici$ cases, $\Tirr$ remains greater than $\Teff$ after this point. In most cases, $\Tirr$ strongly dominates $\Teff$. We consider now how this fact impacts the donor's and binary's evolution.

\subsubsection{Our Irradiation Modelling} 
The general effect of external irradiation is to increase the temperature of 
the donor's atmosphere \citep{milne26}, which will tends to lower $L$ and slow 
its cooling \citep[see, e.g.,][in the context of irradiated planets]{burrows03,
baraffe03}.  The amount of heating depends on many factors: the irradiating flux's intensity, spectrum, anisotropy, and the opacity sources in the donor's atmosphere that determine its albedo \citep[see, e.g.,][]{vaz85,barman01, sudarsky03, barman04,burkert05}.  The detailed modelling of all these effects involves a multidimensional, non-grey, radiation-hydrodynamics problem and is beyond the scope of this paper. Thus, to model irradiation's impact, we continue to assume grey opacities and alter our temperature outer boundary matching condition to $T_e = \Tphot$, where $\Tphot$ is defined by
\begin{equation}
\Tphot^4 = \Teff^4 + \eta \frac{\Tirr^4}{4} = \frac{1}{4 \pi \sigma} \left(\frac{L}{R_2^2} + \eta \frac{\Lacc}{4 a^2} \right)\,,
\label{eq:irr_Tsurf}
\end{equation}
where we assume a point-source geometry for $\Lacc$ and complete redistribution of the irradiating flux around the donor's surface \citep{ritter00}. We also define the quantity $\Lsurf = 4 \pi \sigma R_2^2 \Tphot^4$, which gives the total (intrinsic plus thermalized irradiative) luminosity off the donor's surface. The factor $\eta$ is a dimensionless efficiency parameter giving the fraction of $\Lacc$ that is thermalized in the donor's photosphere.  Equation (\ref{eq:irr_Tsurf}) makes it clear that $\eta$ parametrizes (within a grey, 1D model) our \emph{entire} ignorance associated with the above uncertainties (including uncertainties in the assumed geometry).

For our calculations, we choose a fixed $\eta \leq 1.0$ and 
calculate $\Lacc$ from the system's \emph{secular} $\Mdot$. However, AM CVn binaries 
experience a phase in which instabilities in the He accretion disk produce 
cyclical variations in $\mdot_1$ \citep{tsugawa97}.  While these outbursts produce brightness variations $\lesssim 4$ magnitudes, the outburst period is $\sim 5$ days \citep{wood87,patterson97,patterson00}. This corresponds roughly to a $\tauth$ at $m' \lesssim 10^{-10} M_2$, so almost the entire donor is only aware of the time-averaged $\Lacc$,  making our use of the secular $\mdot$ reasonable.

\subsubsection{Impact of Irradiation on Donor \& Binary Evolution}
A star's outer boundary condition only has a significant impact on its structure when the star is nearly fully convective \citep[see, e.g.,][Sec. 10.3]{kippenhahn90}.  Further, during the adiabatic phase of AM CVn evolution, mass loss, not thermal processes, dominate the donor's evolution.  Thus we can expect that irradiation will affect the donor's evolution most during the cooling phase, by which time the donor's are fully convective and thermal processes dominate their evolution.  This expectation is borne out by our numerical calculations.

\begin{figure}
\plotone{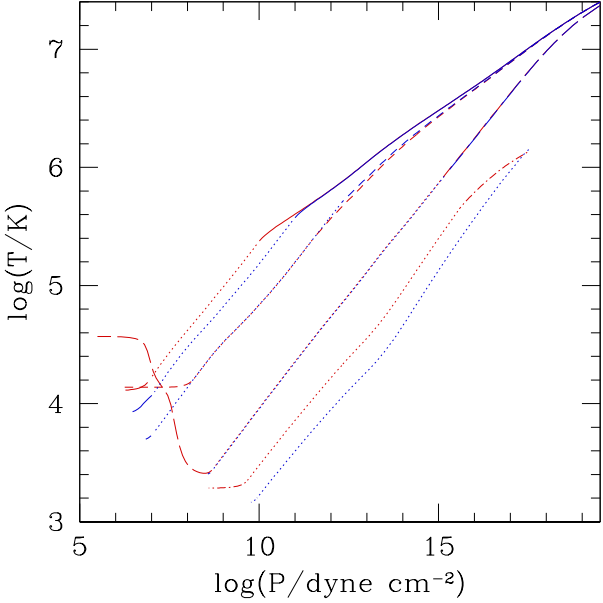}
\caption{The $T(P)$ profiles for an $\eta = 0.5$ irradiated (red lines) and 
non-irradiated (blue lines) donor at points in their evolution when $M_2 = 0.1999$ (solid lines), $0.1998$ (short-dashed), $0.1904$ (dashed), and  $0.015 \msun$ (short-dash dotted); dotted portions of the lines indicate convective regions. The initial conditions for both sets of calculations were $\Mtwi=0.2$, $\Moi= 0.625 \msun$, and $\Rtwi = 0.0380 \rsun$.  Only once the non-irradiated donor is mainly convective and has begun its cooling phase ($M_2 = 0.015 \msun$) do significant interior structural differences with the irradiated model appear.  
}
\label{fig:irr_TPcomp}
\end{figure}

To illustrate this, we show in Figure \ref{fig:irr_TPcomp} a comparison of $T(P)$ profiles between an irradiated ($\eta=0.5$, red lines) and non-irradiated ($\eta=0.0$, blue lines ) donor at several specified values of $M_2$ along respective binary evolution calculations.  Apart from the $\eta$ differences, the initial conditions for the two calculations are the same. The profiles for $M_2 = 0.1999$ (solid lines), $0.1998$ (short-dashed) and $0.1904 \msun$ (dashed) all occur during the turn-on and adiabatic phase. By  $M_2 = 0.015 \msun$ (short-dash dotted line), the non-irradiated donor is in its cooling phase.  

Throughout, $\Tirr$ dominates $\Tphot$, so that the $\eta=0.5$ donor's $\Tphot$-evolution tracks that of $\mdot$. By increasing $\Tphot$, a constant $\Tirr$ will tend to reduce $L$  \citep{baraffe03,burrows03,arras06}. Here, there is at least one other effect important to the $L$ evolution.  Irradiated envelopes tend to have more extensive radiative regions and steeper entropy profiles. During rapid mass loss, this increases the $L$ decrement a mass element experiences as it is advected outwards.

Both effects contribute to lowering the donor's $L$ in the irradiated model. During phases of rapid mass loss, the latter effect even leads to a net $L<0$ (most obviously seen in the inverted $T(P)$ profile of the $M_2 = 0.1904 \msun$ case, but also present at $M_2 = 0.1998 \msun$).  A net $L<0$ is produced when the flux cost required to advect material up the steep entropy gradient cannot be provided by donor's intrinsic flux; the deficit is made up by absorption of irradiating flux and an inverted $T(P)$ profile results.  Although some of the irradiating energy is absorbed below the photosphere, never more than $10^{-6} M_2$ of the donor is involved and the inward directed flux is always $\lesssim 0.01 \eta\, \sigma\, \Tirr^4$. The persistence of the $L<0$ condition depends on $\Mdot$ and $\eta$, with $L$ recovering to positive values once $\Mdot$ decreases sufficiently.  For the case shown in Fig \ref{fig:irr_TPcomp}, $L<0$ until $M_2 \approx 0.02 \msun$, beyond the point the non-irradiated donor started its cooling phase.

Figure \ref{fig:irr_TPcomp} further illustrates our general results that differences between the deep interior structures of irradiated and non-irradiated donors only extend as far as the base of the convective layer, roughly speaking.  Thus it is only once the donors become nearly fully convective do significant structural differences occur; such differences become even more apparent once the non-irradiated donor begins its cooling (e.g., the $M_2 = 0.015 \msun$ profiles in Fig. \ref{fig:irr_TPcomp}).
 
\begin{figure}
\plotone{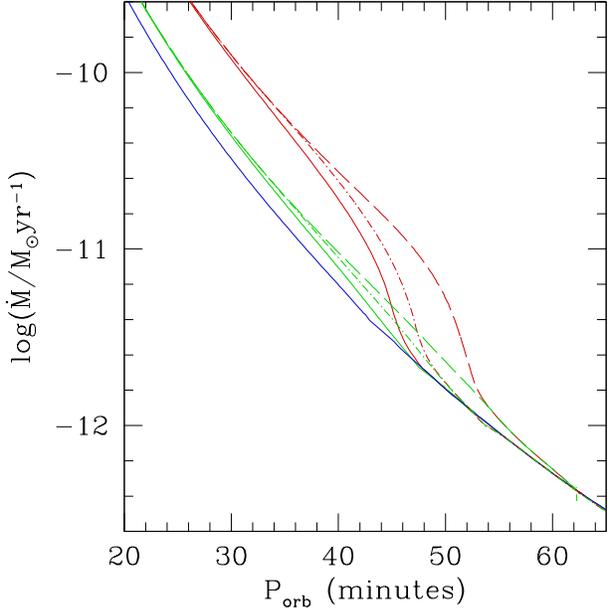}
\caption{The impact of irradiation on the binary's $\Mdot$ evolution for the initial conditions $\Mtwi=0.2$ and $\Moi=0.625 \msun$. Line colour indicates $\log(\psici)$: 1.1 (red), 2.0 (green), and 3.0 (blue); line style indicates $\eta$: 0.0 (solid), 0.1 (short-dash dotted), and 0.5 (dashed).  Only the non-irradiated $\log(\psici)=3.0$ case is shown due to numerical difficulties in converging the irradiated models during the turn-on phase.  However, from other irradiated models with $\log(\psici)=3.0$  the irradiated $\log(\psici)=3.0$ tracks will not differ substantially from the $\eta=0.0$ track shown.  For lower $\psici$, irradiation extends the adiabatic phase and slows the donor's cooling, elevating $\mdot$ during these phases.  %
}
\label{fig:irr_MdPo}
\end{figure}
\begin{figure}
\plotone{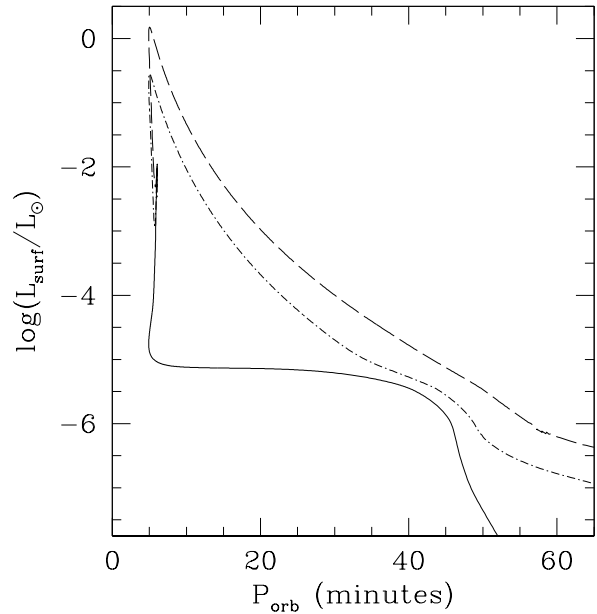}
\caption{Irradiation's impact on the donor's  $\Lsurf$ in the limit 
of zero Bond albedo for systems sharing the initial conditions of $\Mtwi=0.2$, $\Moi=0.3 \msun$, and $\log(\psici)=$ 2.0.  Line style has the same meaning as in Figure 
\ref{fig:irr_MdPo}.  In the $\eta \neq 0$ cases, irradiation substantially 
elevates $\Lsurf$. }
\label{fig:irr_TeffL}
\end{figure}

By decreasing $L$, the net effect of irradiation is to extend the adiabatic phase of evolution to lower $M_2$ and to slow the donor's cooling afterwards. Irradiated donors continue expanding to longer $\Porb$ and contract more slowly once cooling begins.   This 
alters the $\Mdot$ evolution, as illustrated in Figure \ref{fig:irr_MdPo}.  The 
solid lines show systems' evolution with non-irradiated donors, the short-dash 
dotted and dashed lines show $\eta=0.1$ and $\eta=0.5$ tracks, respectively. 
It can be seen that differences in $\Mdot$ evolution between $\eta$ values 
appear even before the cooling phase for non-irradiated case.  This reflects 
the outer boundary condition's growing importance in $\eta=0.0$ donors, which 
have much deeper convective regions by this point in the evolution.  As $\eta$ 
is increased, the $\Porb$ where the $\Mdot$-decline occurs increases; this 
effect is more pronounced at lower $\psici$.  Before contracting to the fully 
degenerate track, irradiated donors converge to an $\eta$-dependent, intermediate 
cooling track.  In Fig. \ref{fig:irr_MdPo}, this occurs at $\Porb \approx 49 (54)$ 
minutes for the $\eta=0.1 (0.5)$ tracks. By this point, the irradiated donors are 
fully convective, so that $R_2$ depends on $s_c$, $M_2$, and the strength of 
irradiation \citep{arras06}.  As irradiation is a function of $M_2(R_2)$ through 
$\Mdot$, the evolution of fully-convective donors within our model are 
parametrized exclusively by $\Mtot$ and $\eta$, as evidenced by this intermediate 
convergence.

The evolution of $\Lsurf$ for a set of donors with differing $\eta$ at fixed $\Moi$, $\Mtwi$, and $\psici$ are shown in Figure \ref{fig:irr_TeffL}. Within our grey-atmosphere modelling, Fig. \ref{fig:irr_TeffL} can be interpreted as providing the donor's surface luminosity in the limit of zero Bond Albedo. Since $\Tirr$ dominates $\Tphot$ over most of the evolution, the $\Lsurf$ evolution mirrors that of $\Mdot$.   During the turn-on phase, $\Lsurf$ decreases initially in all donors.  This decrease is reversed for $\eta \neq 0$ by the $\Mdot$ growth.  By $\Mdot$ maximum, $\Lsurf$ $\approx 10^3$ -$10^5$ higher in irradiated models compared to non-irradiated donors.  During the adiabatic phase, $\Lsurf$ decreases with $\Mdot$, while the non-irradiated donor's $L$ plateaus. In the cooling phase $\Lsurf$ converges towards tracks parametrized by $\eta$, but along much shallower slopes than non-irradiated donors. Even so, by $\Porb \approx 60$ minutes, even $\eta=0.5$ donors are still rather dim, with $\Lsurf \lesssim 10^{-6} \lsun$.
 
\section{Discussion and Applications}
\label{sec:discussion}

\subsection{Comparison to Prior AM CVn Donor Models}
\label{sec:discussion_priormodcomp}
\begin{figure}
\plotone{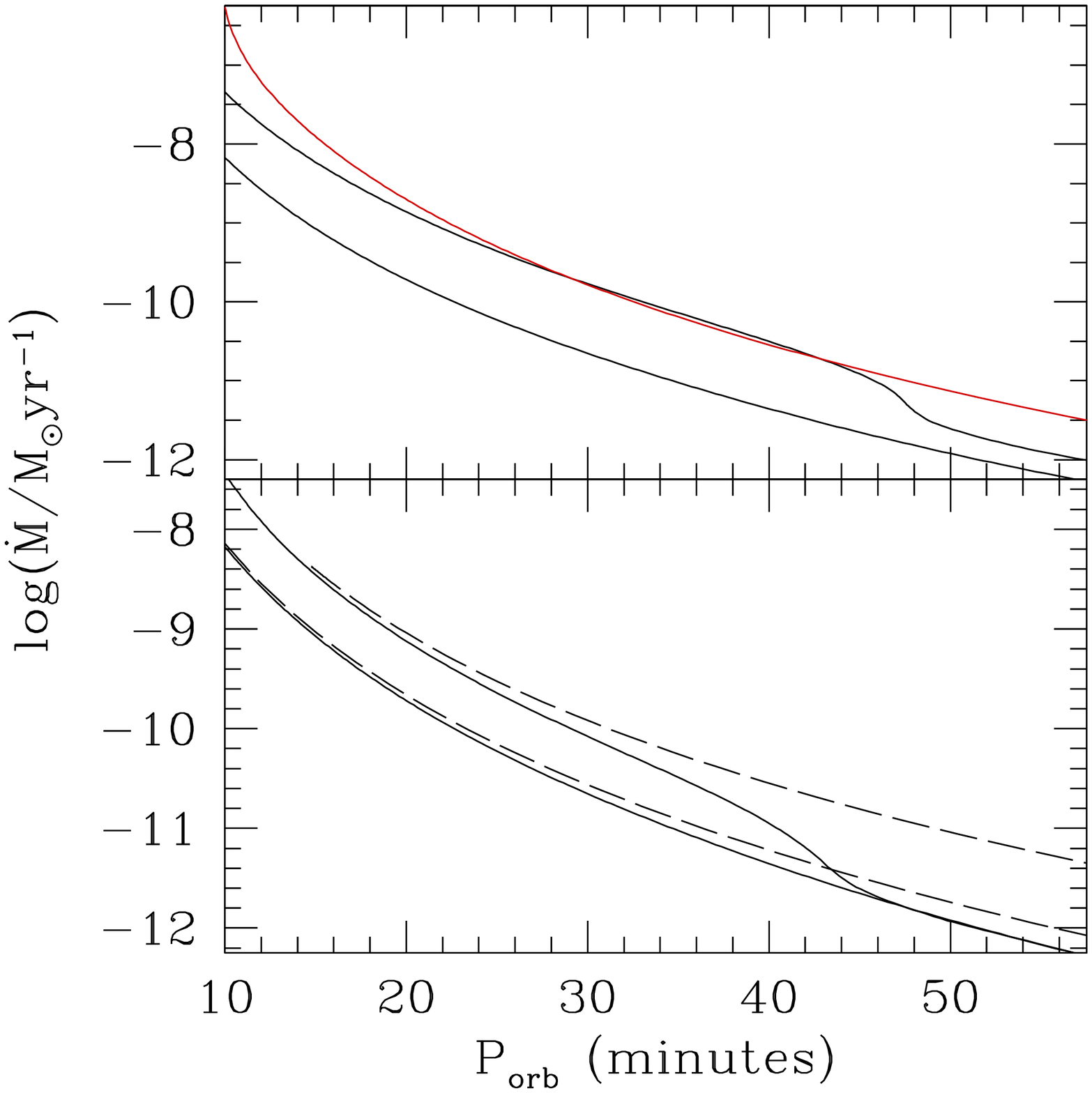}
\caption{Comparison of AM CVn system's outward $\Mdot$-$\Porb$ evolution produced 
by different donor models. The lower panel compares evolution with our current models (solid lines) to that with the \citet{deloye03} isentropic models (dashed lines). The two solid lines had $\Mtwi=0.2 \msun$ and $\log(\psici) = 1.1$ (upper) and $3.5$ (lower); in all four systems $\Mtot = 0.5 \msun$.  The initial conditions ($M_1$, $M_2$, $R_2$) for each dashed line evolution equalled the set of these values on the corresponding solid line at the point of intersection.  The upper panel compares evolution with our current models (solid lines, lower same as in lower panel,  upper with $\Mtot = 1.325$, $\Mtwi = 0.3 \msun$, $\log(\psici)=1.1$) with the evolution produced by the semi-degenerate donor $M$-$R$ relation of \citet{nelemans01a} with $\Moi=0.4$ and $\Mtwi=0.3\msun$ (red line). \label{fig:MdPo_priormod_comp}}
\end{figure}

We start our discussion by comparing the $\Mdot(\Porb)$ evolution produced in our current donor models to that produced by several prior donor models in Figure \ref{fig:MdPo_priormod_comp}.  The lower panel focuses on the comparison between our current models (solid lines) and the \citet{deloye03} isentropic models (dashed lines). The \citet{deloye03} models assume the donors are fully convective and application of these models in both \citet{deloye03} and \citet{deloye05} also assume adiabatic donor evolution. Our more complete, current modelling shows that these He donors are not fully convective over much of the AM CVn evolution phase and that adiabatic evolution only occurs out to $\Porb \approx 40-55$ minutes. 

The impact of these differences is apparent between the $\Mdot$ evolution shown by the dashed and solid  lines in Fig. \ref{fig:MdPo_priormod_comp}. The initial conditions (i.e. $M _1$, $M_2$, $R_2$) for each dashed line evolution equals the set of these values along the corresponding solid line at the point of intersection. From there, the isentropic donors produce $\Mdot$ evolution that increasingly diverges upwards from the solid-line tracks, even during the adiabatic evolution phase. This results from the realistic models being radiative throughout much of their core.  Thus, compared to isentropic donors, the realistic models have a greater  $\nrtw < 0$, even during adiabatic evolution (\S \ref{sec:donorev_preR2min} and Appendix \ref{app:R_response}). The differences between these evolution tracks become even more dramatic once the realistic donors begin cooling. Thus, modelling ultracompact binary evolution using isentropic donors will overestimate the $R_2$ expansion rate, the $\Mdot$-$\Porb$ relations, and $\Porb(t)$.

This has implications for how $\Mdot$ measurements constrain an AM CVn system's 
formation channel.  \citet{deloye05} showed that assuming isentropic donors leads to a significant overlap in the $\Mdot$-$\Porb$ plane between hot WD channel systems and He-star channel systems.  The overlap between these channels is much reduced by with our current donor models, as illustrated in the upper panel of Fig. \ref{fig:MdPo_priormod_comp}.  There the black lines show  the maximum range in the $\Mdot$-$\Porb$ relation for WD channel AM CVn systems produced with our current unirradiated models. The red line shows the \citet{nelemans01a} fit to a semi-degenerate donor $M(R)$ evolution with $\Mtot = 0.7 \msun$, and provides a rough lower limit to He-star channel systems' $\Mdot(\Porb)$ evolution.  Only the most massive WD channel systems have any overlap with He-star channel systems.  Thus, given this current theory, a determination of a secular $\Mdot$ significantly above the upper black Fig. \ref{fig:MdPo_priormod_comp} would indicate a system formed through the He-star channel.  This statement has one caveat: if WD channel systems have a maximum 
entropy greater than indicated by our determinations in \S \ref{sec:initial_models} 
or if the donors are heated significantly (e.g., by tidal mechanisms) earlier in 
the AM CVn phase, then the overlap with the He-star channel systems could be 
increased.

We point out that modelling He-star channel systems with \citet{nelemans01a} 
semi-degenerate $M(R)$ relation beyond $\Porb \gtrsim 45$ is problematic. This is 
because these donors will also begin cooling and contracting by these $\Porb$, 
similar to the WD channel donors.  Thus, in reality, the red line in Fig 
\ref{fig:MdPo_priormod_comp} should begin a down turn somewhere in the vicinity 
of the $\Mtot=1.325 \msun$ track.  In fact, the original calculation used by 
\citet{nelemans01a} to determine this fit \citep[model 1.1 of][see their Fig. 
2]{tutukov89} shows the start of this down turn at $\Porb\approx 40$ 
minutes. Thus, beyond the $\Porb$ at which donors begin their cooling, $\Mdot$ measurements alone will not distinguish between formation channels.

\subsection{The Orbital Period Distribution of WD-Channel AM CVn Binaries}
\label{sec:discussion_Porbdist}
While the changes to the adiabatic phase $M_2(R_2)$ evolution will quantitatively alter our expectations for the AM CVn population's $\Porb$-distribution, the detailed evolution to $\Porb$-minimum and the occurrence of donor cooling will produce qualitatively new features in the population's $\Porb$-distribution. We discuss these features here.

A key factor in how an individual system contributes to the $\Porb$-distribution is the time-derivative, $\Pdot$, of its $\Porb$-evolution.  In a steady-state, continuity requires that the number density of systems at some  $\Porb$, $n_{\Porb}$, scale as $n_{\Porb} \propto |1/\Pdot|$ \citep[see, e.g.,][]{deloye03}.  Here we will use this scaling relation to display steady-state distributions of systems sharing initial data.  This provides a straightforward means for displaying how differing initial conditions influence the relative contribution systems make to the overall $\Porb$-distribution.  We calculate $\Pdot$ along an evolutionary track via
\begin{equation}
\frac{\Pdot}{\Porb} = 3 \left[ \left(\JdotJ\right)_{\mathrm{GW}} - \frac{\mtwodot}{M_2} (1-q) \right]\,.
\label{eq:Pdot}
\end{equation}

\begin{figure}
\plotone{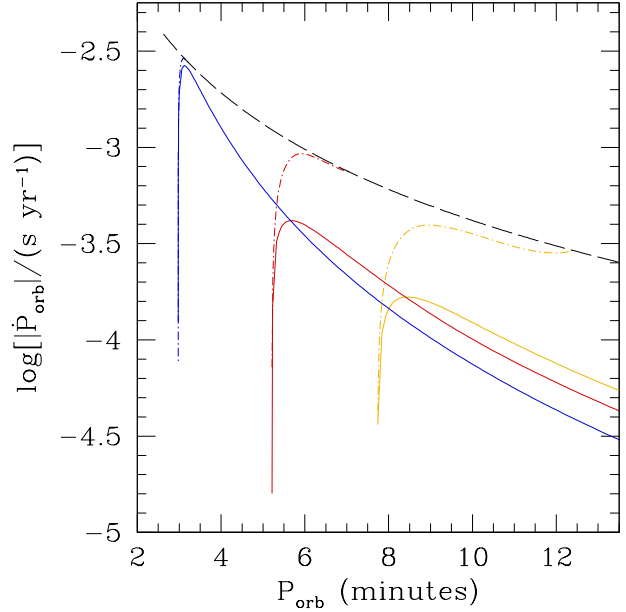}
\caption{The evolution of $\Pdot$ about the $\Porb$-minimum for representative systems with $\Mtwi=0.25 $ and $\Moi = 0.575 \msun$.  Line colour indicates $\log(\psici)$: 1.1 (yellow), 1.5 (red), and 3.0 (blue).  Short-dash dotted line segments indicate $\dot{P} < 0$, solid lines indicate $\dot{P} > 0$.   The black line shows the corresponding inward $\Pdot$ evolution produced by GW emission alone.}
\label{fig:PdotPcontact}
\end{figure}

We show the evolution of $\Pdot$ about the $\Porb$-minimum for a representative set of systems in  Figure \ref{fig:PdotPcontact}. These systems have $\Moi=0.575$, $\Mtwi = 0.25 \msun$, with colours indicating different $\psici$. The short-dash dotted segments indicate $\Pdot < 0$, solid segments $\Pdot > 0$.   The black line shows the inward evolution due only to the $(\dot{J}/J)_{\mathrm{GW}}$ term in equation (\ref{eq:Pdot}) for comparison. As discussed in \S \ref{sec:donorev_preR2min}, the evolution here is most sensitive to $\Mtwi$ and $\psici$.  The $\Pdot$ evolution reflects this, with $\psici$ (as illustrated in this figure) and $\Mtwi$ both affecting the the value of the $\Porb$ minimum and how $\Pdot$ diverges from the GW-only evolution once contact occurs.  The generic features of our this $\Pdot$ evolution are the strong spikes at $\Porb$-minimum as $\Pdot$ evolves through zero and the existence of $\Pdot$ minimum and maximum that occur before and after $\Porb$-minimum, respectively. 

The relative magnitude of $\Pdot$ before and after the $\Porb$-minimum---which determines the relative number of systems evolving inward vs. outward in a steady-state---depends on $\Mtwi$ and $\psici$. The  $\Pdot < 0$ just before $\Pdot$-minimum is approximated by the GW-only rate, which scales as $(\Jdot/J)_{\mathrm{GW}}$. By $\Pdot$-maximum, $\Mdot$ has achieved its secular rate:
\begin{equation}
\left(\frac{\mtwodot}{M_2}\right)_{\mathrm{eq}} = \frac{(\Jdot/J)_{\mathrm{GW}}}{5/6 + \nrtw/2 - q} \,,
\label{eq:mdot_eq}
\end{equation}
producing a $\Pdot$ during expansion:
\begin{equation}
\left(\frac{\Pdot}{\Porb}\right)_{\mathrm{eq}} = 3 \left(\JdotJ\right)_{\mathrm{GW}} \left[ \frac{\nrtw - 1/3}{\nrtw + 5/3 -2 q} \right] \equiv  3 \left(\JdotJ\right)_{\mathrm{GW}} \beta_{\dot{P}}\,.
\label{eq:Pdot_eq}
\end{equation}

If $M_2$ does not change appreciably from $\Pdot$-minimum to $\Pdot$-maximum, the term in square brackets, denoted here by $\beta_{\dot{P}}$, estimates the relative magnitudes of $\Pdot$ before and after $\Porb$-minimum. Note that $\beta_{\dot{P}}$ depends on $\psici$ through $\nrtw$.  During the early expansion phase, $\nrtw$ evolves from $0$ to $\approx -0.3$ to $-0.05$.  With $0 \leq q \leq 2/3$ (the $q$ range considered here),  $\beta_{\dot{P}}$ can vary considerably.  Typically, $-1 < \beta_{\dot{P}}<0 $, as seen in Fig. \ref{fig:PdotPcontact}.  However, for larger $q$ and $\psici$, $\beta_{\dot{P}}$ can be much less than -1; we see this behaviour in our calculations with  $\Moi=0.3$, $\Mtwi=0.2 \msun$ for  $\log(\psici) \geq 3.0$.  Lower $\psici$ donors have smaller relative outward-to-inward $|\Pdot|$, for two reasons.  One, lower $\psici$ leads to larger minimum $\nrtw$, producing slower expansion and $\beta_{\dot{P}}$ values closer to zero.  Two, $M_2$ during the $\Pdot$-transition is not fixed, with $M_2$ losses greater for lower $\psici$.  This results in a lower $\Jdot$ loss rate post $\Pdot$-maximum, further contributing to smaller $\Pdot >0$.

\begin{figure}
\plotone{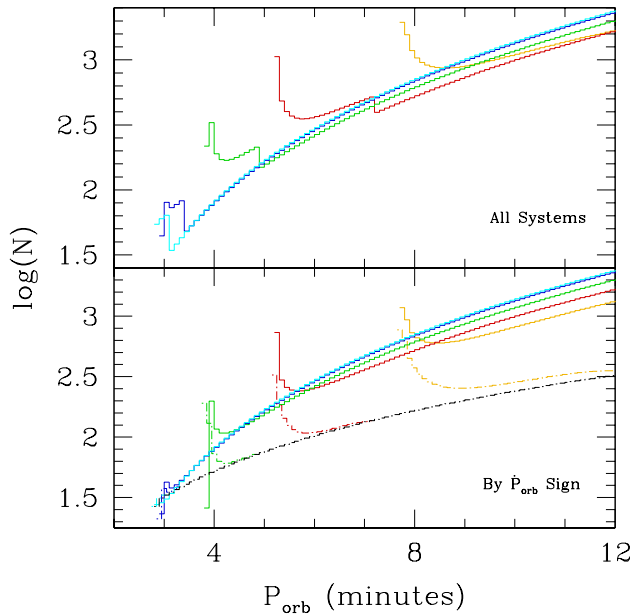}
\caption{Histograms of relative numbers of AM CVn systems in a steady-state along single evolution tracks. All tracks have $\Moi=0.575$ and $\Mtwi=0.25 \msun$ with colour indicating $\log(\psici)$: 1.1 (yellow), 1.5 (red), 2.0 (green), 3.0 (blue), and 3.5 (cyan).  The lower panel separates systems by $\Pdot$ sign: the short-dash dotted histograms counts inward moving systems, the solid histograms outward moving systems.  The upper panel displays the sum of both inward and outward moving systems. The black short-dash dotted line shows the  distribution for the evolution produced by GW emission alone (i.e. $\Mdot=0$). The bins have width of $\Delta \Porb = 6$ s. The overall normalization is arbitrary, but the relative normalization accurately reflects the relative $\Pdot$ rates. }
\label{fig:Nhist_400}
\end{figure}

How the evolution near $\Porb$-minimum translates into the steady-state $\Porb$-distributions is shown in Figure \ref{fig:Nhist_400}.  There we display histograms of relative system numbers along evolutionary tracks with  $\Moi=0.575$, $\Mtwi=0.25$.  The number in each bin, $N$, is calculated by integrating $n_{\Porb} d\Porb$ over each bin.  The overall normalization is arbitrary but fixed across tracks, so that the histograms accurately reflect relative $\Pdot$ rates. The lower panel displays separately the contributions from the $\Pdot <0$ (short-dash dotted lines) and $\Pdot > 0$ (solid lines) segments of the evolution.  The black line in the lower-panel shows the corresponding result for a GW-only driven in-spiral.  The upper panel shows the sum of contributions from both inward and outward evolving systems.

The general trends in the lower panel are the slight deviation from  GW-only evolution from contact inward, followed by a peak as $\Pdot \rightarrow 0$ at $\Porb$-minimum.  The outwardly evolving systems likewise show peaks just post $\Porb$-minimum.  For larger $\Porb$, lower-$\psici$ tracks lead to fewer systems at a given $\Porb$ since hotter donors have larger $M_2$, producing larger $(\Jdot/J)_{\mathrm{GW}}$  \citep[][see also Fig. \ref{fig:PdotPcontact}]{deloye05}. The sharp steps in the upper-panel histograms result from starting our calculations at the point of contact (i.e. pre-contact evolution is not included), so the location and size of each step is somewhat artificial.  However, since they result from having a definite starting point for the GW-driven in-spiral, as is provided by the CE-event forming these systems, there is a physical basis for expecting their existence.

The most striking feature of these histograms are the strong peaks at each system's $\Porb$-minimum.  The impact of these features on the integrated AM CVn $\Porb$-distribution below $\Porb \approx 15 $ minutes will depend on the distribution of initial conditions, the survival of systems at contact \citep[see, e.g.,][]{marsh04}, and how He-star channel systems contribute in this $\Porb$ range.  Determining how the integrated $\Porb$-distribution depends on initial conditions and the physics determining the outcomes at contact is the subject of current work.  This will be most relevant to future space-based GW-interferometers, such as \textit{LISA}, which will provide a rather complete census of the galactic AM CVn population at these $\Porb$ \citep{nelemans01c} and offer direct observational tests of these predictions.

In \citet{deloye06}, we considered the relevance of the $\Pdot < 0$ phase to the short period X-ray variables, RX J0806+1527 \citep[321 s;][]
{beuermann99} and RX J1914+2456 \citep[569 s;][]{haberl95}. There is still 
some question as to the nature of these two sources \citep[see discussion in]
[and references therein] {deloye06}. If both system's periods are orbital 
and their measured $\Pdot$ \citep{strohmayer04,strohmayer05} secular, then \citet{deloye06} showed that both are consistent with being members of the AM CVn population.  If this is the case, there is the question of whether observing either system in its 
present state is an extremely unlikely event.  

\begin{figure}
\plotone{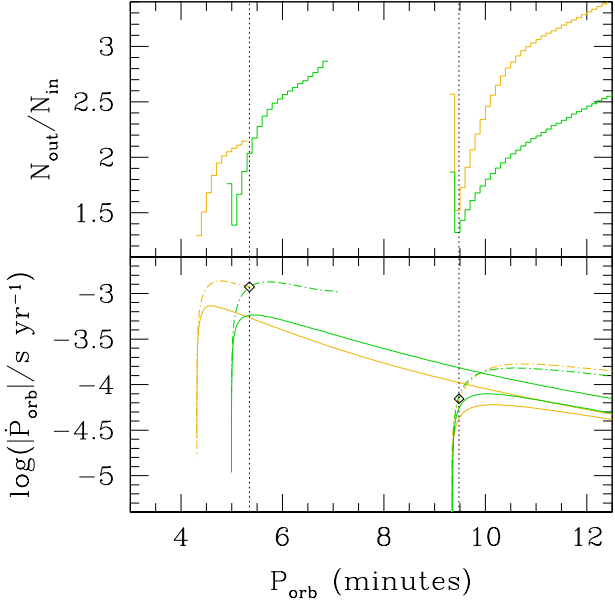}
\caption{The evolution rates for AM CVn system models consistent with measured 
properties of RX J0806+1527 (lines at $\Porb  \approx 5$ minutes) and RX J1914+2456 
(lines at $\Porb \approx 9$ minutes).  The set of two lines shown for each 
system indicates the approximate range of variation allowed by observations 
and our modelling.  Different colours are meant only to guide the eye in matching 
tracks within and between the two panels. The lower panel shows the $\Pdot$ 
evolution for the selected tracks, with solid lines indicating $\Pdot > 0$ and 
short-dash dotted lines $\Pdot < 0$.  The upper panel shows the ratio of  
of number of systems with $\Pdot>0$ to those with $\Pdot <0$ assuming steady-state.  The dotted lines indicates the measured $\Porb$ of each system, while the diamonds show their measured $\Pdot$.  In making these comparisons, we explicitly assume the measured 
$\Pdot$ reflects the system's \emph{secular} $\Pdot$. }
\label{fig:RXsys_PdotP}
\end{figure}

A detailed answer to this question is beyond the scope of this paper. However, a related question can be posed: what is the relative likelihood of detecting each system in its $\Pdot < 0$ versus $\Pdot >0$ phase?  Figure \ref{fig:RXsys_PdotP} provides a partial answer.  The lower panel shows the $\Pdot$ evolution for two sets of AM CVn systems, each of which bracket the evolution of systems consistent with either RX J0806+1527 or RX J1914+2456 \citep[see][]{deloye06}. The upper panel shows, for these evolution tracks, the ratio of the number of systems with $\Pdot>0$ to those with $\Pdot <0$ in steady-state. This ratio is only $\approx 1.3-2.3$ for periods below each system's measure $\Porb$.  Thus, there is not a strong a priori bias to detecting such systems in their $\Pdot >0$ phase, although the actual relative detection likelihood between the two phases is likely influenced strongly by selection effects.  

\begin{figure}
\plotone{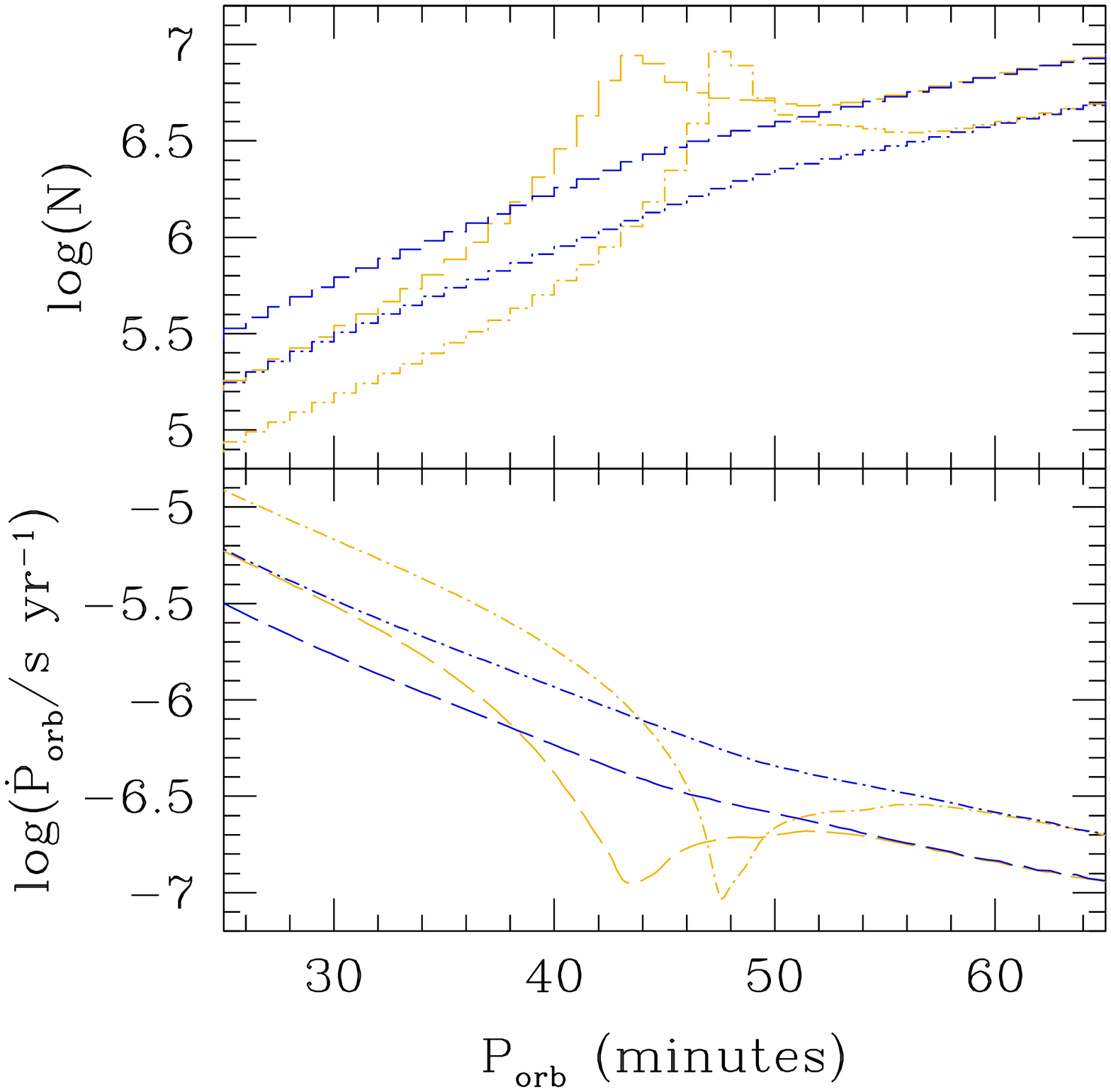}
\caption{The evolution of $\Pdot$ for $\Porb > 25 $ min during the expansion 
phase.  The lower panel shows $\Pdot$ vs. $\Porb$ for two sets of $\Mtot$ each with two different $\log(\psici)$.  Colours indicate $\log(\psici)$ and have the same meaning as in Fig. \ref{fig:Nhist_400}.  Dashed lines show systems with $\Moi=0.35$, $\Mtwi=0.15 \msun$; short-dash dotted lines systems with $\Moi=1.025$, $\Mtwi=0.3 \msun$.  The upper panel shows the relative number of systems expected in steady-state along each evolution track.  As in Fig. \ref{fig:Nhist_400}, the overall normalization is arbitrary, but the relative normalization between tracks is set by their $\Pdot$ rates. The slowing of $\Pdot$ during the donor's cooling phase leads to a peak in system numbers. The peak's $\Porb$ is diagnostic of $\Mtot$ (as well as $\eta$, although this is not show here).  The magnitude of the peak increases with decreasing $\psici$.}
\label{fig:PdotP_Nhist_outward}
\end{figure}

We now turn to how donor cooling influences the $\Porb$-distribution.  By the time $\Pdot$ has passed its maximum, it is evolving at a rate given by equation (\ref{eq:Pdot_eq}), and is essentially determined by $\Mdot$ (which is a function of $\Mtot$ and $\psici$). This is seen in the lower panel of Figure \ref{fig:PdotP_Nhist_outward} where we display $\Pdot$ versus $\Porb$ for two different $\Mtot$ at two different $\psici$.  Before systems begin their cooling phase, the ordering of tracks is for lower $\psici$ donors to produce higher $\Pdot$.  After the cooling phase ends, tracks are distinguished only by $\Mtot$.  During cooling, the donor's contraction stalls the $\Porb$ evolution, producing a reversal in the ordering of $\Pdot$ with $\psici$ and a distinctive peak in the steady-state $\Porb$ distribution along each track (upper panel of Fig. \ref{fig:PdotP_Nhist_outward}).

The location and magnitude of these peaks are determined by $\Mtot$ and $\psici$. 
The system's $\Mtot$ determines the $\Porb$ at which cooling becomes important, 
with larger $\Mtot$ moving the peak's centre to longer $\Porb$.  The donor's 
$\psici$ determines the degree of donor contraction;  hotter donors lead to slower $\Pdot$ during the cooling phase and larger peaks. In slowing donor cooling, irradiation also acts to  shift the peak centres to longer $\Porb$.  The distribution of AM CVns above $\Porb \approx 40$ minutes will provide an integrated diagnostic of the distribution of $\psici$, $\Mtot$, and $\eta$  In this $\Porb$ range, AM CVn binaries will not be 
individually resolvable GW-sources due to the galactic foreground of detached 
WD-WD binaries \citep{nelemans01c}. Thus, whether the system distribution in this $\Porb$ range can be a practical diagnostic tool must await observations progress in the optical/IR wave bands. 

\subsection{The Eclipsing AM CVn SDSS J0926+3624: Evidence for Non-Zero Entropy Donors from the WD Channel}
\label{sec:discussion_hot_donors}
Although we reserve detailed applications of the models presented here to specific 
AM CVn systems for a later companion paper, we shall discuss the recently discovered 
eclipsing AM CVn binary, SDSS J0926+3624 \citep{anderson05} as a system for which 
the entropy of the donor can be probed.  This is especially important as the 
distribution of donor entropy provides (i) a potential diagnostic of AM CVn 
formation channels and (ii) constrains the stellar binary evolution determining AM CVn 
initial conditions \citep{deloye05}. \citet{bildsten06} have presented evidence 
based on accretor properties that the two known long-period AM CVn binaries, 
GP Com ($\Porb = 46.6$ minutes) and CE-315 ($\Porb = 65.1$ minutes) both harbour 
a relatively hot donor.  Additionally, if early-contact AM CVn binaries are 
indeed the correct model for the systems RX J0806+1527 and RX J1914+2456, then 
this would provide further evidence for hot donors: the bracketing models shown 
in Fig. \ref{fig:RXsys_PdotP}, have a range of $\log(\psici) \approx 1.3-1.8$ 
for RX J0806+1527 and $\approx 1.2-1.5$ for RX J1914+2456. SDSS J0926+3624, however, has 
finally provided direct evidence for hot donors in AM CVn binaries. 

This evidence comes in the form of $M_1$ and $M_2$ determinations made via modelling 
of eclipse light-curves \citep{marsh06}. Assuming the accretor obeys a fully degenerate WD $M$-$R$ relation (a good approximation at the determined $M_1$), these authors determine $M_1 = 0.84 \pm 0.05 \msun$ and $q=0.035 \pm 0.002$, giving $M_2 = 0.029 \pm 0.002 \msun$, $\approx 50$\% more massive than a zero-temperature WD that fills its Roche lobe at this system's $\Porb=28.3$ minutes.  This $M_2$ measurement provides us the first direct means of determining an AM CVn donor's current entropy. From the determined $M_2$-range, we find that $\log(\psici)$ lies approximately in the range 1.60--1.35.  If the actual range of He-star channel donor $M$-$R$ evolution does not differ significantly from the \citet{nelemans01a} fit for these systems, then a He-star channel system should have $M_2 \approx 0.05 \msun$ at $\Porb = 28.3$ minutes. The evolve-MS channel also appears to produce a value of $M_2$ which is too high for systems at this $\Porb$ \citep[see Table 1 of][]{podsiadlowski03}.  \emph{Thus SDSS J0926+3624 presents evidence that the WD channel indeed contributes to the observed AM CVn population and that this channel produces non-zero entropy donors as predicted by \citet{deloye05}}.

We can also predict the current $\Mdot$ in SDSS J0926+3624. From our calculations, a system with the determined value of $\Mtot = 0.869 \msun$ and $M_2 = 0.029 \msun$ at $\Porb = 28.3$ minutes, has an $\Mdot = \ee{9.8}{-11} \msun$ yr$^{-1}$.
At fixed $\Porb$ when $M_2 \ll M_1$, $\Mdot \propto M_1^{2/3} M_2^2 \propto M_1^{8/3} q^2$ \citep{deloye05}, so the error bars quoted in \citet{marsh06} for $M_1$ and $q$ provide an error of $\approx 20\%$ on $\Mdot$. Thus, we estimate $\Mdot \approx \ee{9.8}{-11} \pm \ee{2.0}{-11} \msun$ yr$^{-1}$. This $\mdot$ range is close to value at which the accretor's thermal evolution decouples from the compressional heating provided by the accretion \citep{bildsten06}.  Thus, we can use Fig. 1 from \citet{bildsten06} to estimate the accretor's $T_c$ from $\mdot$ and find $T_c \approx 1.8$-$2.1 \times 10^7$ K.  Using these $\mdot$ and $T_c$ ranges, we sum the accretor's cooling and compressional luminosity \citep[see \S 2 of][]{bildsten06} to estimate the accretor's $\Teff \approx 21,300$-$23,800$ K, taking $M_1 = 0.84 \msun$ to determine $R_1 \approx 0.01 \rsun$. More refined estimates of the accretor's thermal properties will require more detailed calculations taking into account variations in $\Mdot$ evolution histories, the time-dependent evolution of the accretor's envelope during the decoupling phase, and possibly the effect of He shell flashes on the accretor's surface.

The measurements of $\Mtot$ and $\psici$ in additional systems are required before 
the observational distributions of these parameters can be determined. However, the 
likelihood of discovering a system such as SDSS J0926+3624 given our theoretical 
models of the WD channel AM CVn population can be considered. We first compare $\Mtot$ 
to the distribution in Fig. 1 of \citet{nelemans01a}.  The locus of points defining 
the region $\Mtot = 0.869 \pm 0.05 \msun$ in this figure lies outside the $M_1$-$M_2$ parameter space these authors consider most likely.  This raises the question of whether observing such a high $\Mtot$ in SDSS J0926-3624 simply is the result of small number statistics or whether this is a hint of additional physics that skews the $\Mtot$ distribution to higher values.  An example of such physics would be the preferential survival of high $\Moi$ systems during a direct impact accretion phase at  contact \citep{marsh04}.  Finally, from Figure \ref{fig:psici_hist}, about 12\% of WD channel AM CVn systems in our modelling have $\psici$ in the range consistent with SDSS J0926+3624, not taking into account the system's high $\Mtot$. Given the rather flat $\psici$ distribution expected from theory, determining $\Mtot$ and $\psic$ constraints in additional AM CVn systems is essential if we are to determine if this is consistent with our theoretical expectations or not.

\subsection{Observational Signatures of AM CVn Donors}
\label{sec:discussion_donor_obs}
\begin{figure}
\plotone{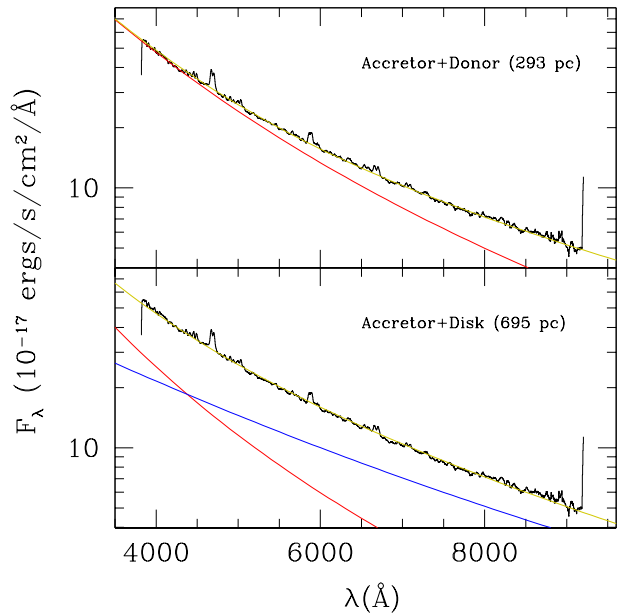}
\caption{Comparison between the Sloan spectrum for SDSS J0926+3624 (black line) 
and theoretical models.  The upper panel shows a model spectrum (yellow line) 
calculated using contributions from a $\Teff=21,000$ K, $R_1=0.01 R_\odot$ 
accretor (red line) and $\Teff=4400$ K, $R_2=0.043 R_\odot$ donor (which 
lies below the x-axis) at a distance of 293 pc.  The lower panel shows the 
combined spectrum (yellow line)  from a $\Teff = 39,000$ K accretor (red line) 
and a steady-state $\alpha$-disk model (blue line) with $\Mdot=9.8\times^{-11} 
\msun$ yr$^{-1}$ at a distance of 695 pc.}
\label{fig:J0926_spectra}
\end{figure}

Here we consider predictions for the donor's  contribution to the system's light in  SDSS J0926+3624 based on our current models. The near edge-on inclination of SDSS J0926+3624 makes this system a good candidate for discriminating the accretor's and donor's light without significant contamination from the accretion disk.  The discovery spectrum of SDSS J0926+3624 \citep{anderson05} is reproduced in Figure \ref{fig:J0926_spectra} from the SDSS archival data. It was obtained over a 3600s exposure---roughly two system orbits---and thus provides a phase averaged spectrum. It is decidedly non-blackbody in shape, indicating that a DB WD accretor is not the only contributor to the system's light.

The question is whether the second component contributing to the flux is the donor or the disk? The \citet{marsh06} $M_1$, $M_2$ constraints imply $R_1 \approx 0.01 R_\odot$ and $R_2 = 0.043 R_\odot$. From our current modelling, we predict the donor's $\Teff \approx 1750$-$4500$ K, corresponding to an $\eta$ range of $0.0$-$0.5$.  With  $i = 83.1\degr$ and $q=0.035$  \citep{marsh06}, the projected area of the disk's face $\approx 2.7 R_2^2$; i.e., out of eclipse roughly equivalent areas of disk and donor surfaces are seen. Thus, the disk may provide a significant contribution to the system's light. Indeed, \citet{marsh06} find that the disk contributes $\approx 50$\% of the flux in the $r'$ band and $\approx 25$\% in the $g'$.  They also find significant inter-orbit variability, so whether this particular ratio of flux contributions is representative of conditions during the Sloan observation is not certain.

Given the prominent double-peaked He emission lines in this system, AM CVn phenomenology would argue the disk is either in a stable, seemingly optically thin,  low-state or is an outbursting disk caught in quiescence. In either case, a stable, optically thick $\alpha$-disk model spectra is not expected.  To check this, we calculated an $\alpha$-disk spectra assuming $\Mdot=9.8\times^{-11} \msun$ yr$^{-1}$ and an outer radius of $0.7 a$.  We added this to the accretor's flux modelled as a single $\Teff$ blackbody.  We then adjusted this $\Teff$ and the system's distance to give a ``by eye'' best fit to the SDSS J0926+3624 spectrum.  The results are shown in the lower panel of Figure \ref{fig:J0926_spectra}.  A rather hot, $\Teff = 39,000$ K, accretor and a system distance of 695 pc is required for a reasonable fit.  This model underestimates the continuum flux at wavelengths $\lambda \lesssim 4000$ \AA and $\lambda \gtrsim  8000$ \AA.  Additionally, the $r'$ flux from the disk is significantly greater than the accretor, in disagreement with \citet{marsh06}.  This however may not be a significant issue given the source's variability.

We also considered a model with accretor and donor contributions, but no disk.  Both components were modelled as single $\Teff$ blackbodies.  The observed spectra is well fit by the combination of a $\Teff=21,000$ K accretor and a $\Teff = 4,400$ K donor at a distance of 293 pc (upper panel of Figure \ref{fig:J0926_spectra}).  The accretor's $\Teff$ is a little below the range predicted by the \citet{bildsten06} theory (\S \ref{sec:discussion_hot_donors}), while the donor's $\Teff$ is consistent with an $\eta$ somewhat below 0.5. The fit in this case is somewhat better than the accretor+disk model. We should note that other acceptable fits can be found by collectively increasing or decreasing both components' temperatures and the system's distance.  Thus, higher accretor $\Teff$ require higher $\eta$ values for the donor to be hot enough to produce a good fit to the spectra.

A parallax measurement for SDSS J0926+3624 will clearly distinguish between the stable $\alpha$-disk+accretor and the accretor+donor model (as well as further constrain the component temperatures in this latter model).  The better agreement between data and donor+accretor model and our expectation that the disk is not in a stable high-state already argues for the donor+accretor model.  Given our poor understanding of a low-state disk spectra and flux, however, one can only conclude that an accretor+donor model is fully consistent with both the observed spectrum and all theoretical expectations for SDSS J0926+3624.  Since this model doesn't include any disk contributions, it does fail to explain the inferred disk properties of the \citet{marsh06} results.  It may be possible that it is a cool disk providing the long-$\lambda$ flux in this system,  possibly explaining both the Sloan spectrum and the \citet{marsh06} results.  Better understanding of quiescent He disks and phase-resolved spectral studies of this system may both be required to break this degeneracy between either a cold disk or donor as the source of long wavelength flux.

\section{Summary}
\label{sec:summary}
We have implemented a new stellar evolution code in C++ that allows significant flexibility in use of input physics and defining systems of ODEs. This code has been used to model, for the first time, the full stellar structure of the donors in WD-channel AM CVn binaries. Specifically, the thermal and structural evolution of the donor has been calculated to determine how the donor's thermal evolution affects the evolution of the binary and to provide the first predictions for the donor's light in these systems.

Systems forming through the WD channel are expected to have a range of donor properties, most importantly a range of initial (i.e., at contact) degeneracy \citep{deloye05}.  We modelled the pre-contact evolution of WD channel systems based on the \citet{nelemans01a} population synthesis to determine suitable initial donor models for our subsequent calculations.  The donors in this population have initial central degeneracy parameters between $\psici \approx $10-$10^4$, with the distribution in $\psici$ being rather flat between $\psici \approx $25-4000.  Most of these systems make contact at $\Porb \approx $2-11 minutes. This range of donor parameters is likely dependent on assumptions about the binary evolution leading to their formation (in particular the CE-event prior to the AM CVn phase) and our assumptions about the donor's cooling during the pre-contact phase.  Quantifying the potential effects of these assumptions will require more extensive modelling of the proto-donors in this pre-contact phase.

Our evolutionary calculations show that WD-channel AM CVn systems have three phases of evolution.  An $\Mdot$ turn-on phase, during which $\Mdot$ grows to its maximum value while $R_2$ contracts under mass loss, and $\Porb$ decreases.  This behaviour produces a turn-on phase lasting significantly longer than previous estimates \citep[e.g.,][]{marsh05,willems05}: $\sim 10^4$-$10^6$ yrs depending on the donor's initial $\psici$. In the second phase, the donor expands adiabatically under mass loss.  The third phase begins once the mass loss rate and the donor's thermal time have have both decreased enough to allow, starting at $\Porb \approx 45$ minutes, the donor to cool and contract to a fully degenerate configuration.  We discussed how the system's initial conditions influence the later evolution of $R_2(M_2)$ and $\Mdot(\Porb)$.  We also revised the upper limit to the $\Mdot-\Porb$ relation for WD-channel systems (given our initial condition determinations). Finally, we predicted the donor's intrinsic $L$ and $\Teff$; during the adiabatic phase, $L \approx $10$^{-6}$--10$^{-4}$ $\lsun$, while $\Teff \approx $1000--1800 K.

The flux generated by the accretion flow in these systems can easily dominate the donor's intrinsic thermal output. We self-consistently modelled the impact of the accretion light on the donor's and binary's evolution in the grey approximation.  The irradiation reduces the donor's intrinsic $L$, producing a delay in the onset of cooling and slowing the donor's contraction once cooling does begin.  This shifts the downturn in the $\Mdot(\Porb)$ relation seen in the non-irradiated donors to longer $\Porb$.  Irradiation also elevates the donor's photospheric temperature and surface luminosity (up to 10,000 K and $10^{-2} \lsun$, respectively, during the adiabatic phase).  The observational signatures of irradiation depend on the irradiating flux's spectra, the opacity sources in the donors atmosphere and the efficiency of day-to-night side energy flow, all considerations beyond this work's scope.  Our predictions for the irradiated donor's light are thus approximate and valid in the limit of complete redistribution of flux in a grey atmosphere with a Bond albedo of zero.

In comparison to prior predictions using simpler donor models,  we find that previous assumptions of fully-convective donors and adiabatic evolution \citep{deloye03,deloye05} are not valid over much of the AM CVn evolution.  The donors only become fully convective at $\Porb \approx 40$ minutes and  $M_2 \approx 0.01$.  Prior to this, their shallower entropy profile leads to a slower expansion rate relative to fully convective models.  Due to this, we argue that prior to the onset of cooling, $\Mdot$ measurements could distinguish between systems formed in the He-star versus WD channels \citep[as opposed to the conclusions of][]{deloye05}.  After the cooling phase develops, all initial donor entropy information, including any that distinguishes formation channels, will be erased.

A system's evolution about its $\Porb$-minimum  and the occurrence of donor cooling phase both leave diagnostic signatures on the AM CVn population's $\Porb$-distribution.  Most significantly, the evolution of $\Pdot \rightarrow 0$ at the $\Porb$-minimum and the slowing of $\Porb$ evolution that occurs during the cooling phase lead to peaks in the $\Porb$-distribution whose location and size depend on initial parameters (and $\eta$ in the case of the cooling peaks).  These could provide observational diagnostics of the distribution of these parameters in the galactic AM CVn population.

Finally, we showed that recent measurements of $M_1$ and $M_2$  in the eclipsing AM CVn system, SDSS J0926+3624 \citep{marsh06} provide direct evidence that WD-channel systems contribute to the AM CVn population and that this channel produces non-zero entropy donors as predicted by \citet{deloye05}.  We compared predictions for this system's light based on our models, showing that a composite spectrum consisting of donor and accretor contributions is fully consistent with this system's discovery spectra.  Based on this, we predict this system lies at a distance of $\approx 290$ pc. 

Current investigations include applying this theory to interpret the improving observational constraints in many of the known AM CVn systems (e.g., Roelofs et al. 2007, submitted). In addition, we are considering how the physics relevant to the AM CVn system formation and early contact-phase survival could be probed by future \textit{LISA} observations of the $\Porb$-distribution of sources at $\Porb \lesssim 15$ minutes.  Other indicated work includes proper non-grey, phase dependent modelling of the donor's irradiated atmosphere as well as progress on understanding the contribution of low-state He accretion disks to the system's emission. 

\section*{Acknowledgments}
We thank the anonymous referee for a careful reading of this manuscript and for suggestions that improved its overall presentation. We thank Susana Barros, Gijs Roelofs, Tom Marsh, Gijs Nelemans, Bart Willems, and Dean Townsley for helpful discussions and encouragement during the preparation of this manuscript, as well as Jason Alexander for providing the low-temperature opacity tables needed in this work and for answering associated questions.  For material support, CJD thanks Craig Heinke, Vicky Kalogera, and Greg and Lauree Hickok, without whose help this work could not have been completed.  This work was supported by NASA through XMM grant NNX06AH62G and by the NSF through grants AST-0200876 and AST-0449558.
\appendix
\section{The $R_2$ Response to Mass Loss}
\label{app:R_response}

The donor's $R_2$ can be expressed as the integral of equation (\ref{eq:drdm}):
\begin{equation}
R_2^3 = \frac{3}{4 \pi} \int_0^{M_2} \frac{dm}{\rho}\,.
\label{eq:rdmint}
\end{equation}
The derivative of equation (\ref{eq:rdmint}) with respect to $M_2$ is
\begin{equation} 
\begin{split}
\der{R_2}{M_2} &= \frac{1}{4 \pi R^2} \left[ \frac{1}{\rhophot} -  \int_0^{M_2} \frac{dm}{\rho} \left(\der{\ln \rho}{M_2}\right)_m \right]\\
&=  \frac{1}{4 \pi R_2^2} \left[ \frac{1}{\rhophot} -  \int_0^{M_2} \frac{dm}{\rho} \left(\frac{1-\chi_T \nabla'}{\chi_{\rho}} \right) \left(\der{\ln P}{M_2}\right)_m  \right]\,, \label{eq:drdmint}
\end{split}
\end{equation}
where $\chi_{\rho} = ( \partial \ln P/\partial \ln \rho)_T$, $\chi_T = (\partial \ln P/\partial \ln T)_{\rho}$ and $\del' = (\partial \ln T/\partial \ln P)_m$ gives the change in $T$ at fixed $m$ that occurs due to changes in $M_2$.

The quantity $(d \ln P/dM_2)_m \sim d\ln P/dm \sim P_c/(P M_2)$, showing the donor's outer layers dominate the $R_2$ response to $M_2$. Also, $ (\partial \ln T/\partial \ln P)_m > 0$ and  $\nabla' \leq \delad$ generically in our donors, so the local response to mass loss at fixed $m$ is an expansion in $r$.  The surface term then is the only driver of $R$ decrease under mass loss \citep[see also][]{hjellming89}. Note that both these behaviours are apparent in Fig. \ref{fig:R_lagrange_ev}.

How the $R_2$ evolution depends on the donor's structure is more easily seen by transforming to $P$-coordinates and considering only the donor's outer layers. The contribution, $\delta R_2$, to $R_2$ between surface at $\Pphot$ and some pressure $P_b \gg \Pphot$:
\begin{equation}
\delta R_2 =  \frac{1}{g_b}  \int_{\Pphot}^{P_b} \frac{dP}{\rho}\,.
\end{equation}
where $g_b \approx$ const. is the gravitational acceleration at $P_b$.
The change in this layer's thickness under mass loss is then:
\begin{equation}
\der{(\delta R_2)}{M_2} = - \frac{1}{g_b} \int_{\Pphot}^{P_b} \frac{dP}{\rho} \left( \der{\ln \rho}{M_2}\right)_P \,,
\end{equation}
where we have neglected for simplicity the surface term's contribution.  The layer's response the depends on $(d\ln\rho/d M_2)_P$, which can be rewritten as
\begin{equation}
\begin{split}
 \left( \der{\ln \rho}{M_2}\right)_P &= \left[  - \tder{\ln \rho}{\ln P}{M_2} + \tder{\ln \rho}{\ln P}{m} \right] \left( \der{\ln P}{M_2}\right)_m\\
&=  \frac{\chi_T}{\chi_{\rho}} \left( \nabla - \nabla' \right) \left( \der{\ln P}{M_2}\right)_m\,. 
\label{eq:drdm_Pint}
\end{split}
\end{equation}

\section{The Luminosity Profile's Response to  Mass Loss}
\label{app:L_response}
To examine how the donor's $l$-profile evolution in response to mass loss depends on the donor's structure and relative ordering of $\tauth$ and $\taum$, taking $\epsilon=0$ we can rewrite equation (\ref{eq:dldm}) as
\begin{equation}
\begin{split}
\der{\ln l}{m}  &= - \frac{c_P T}{l} \left( \nabla' - \delad \right) \tder{\ln P}{t}{m}\\
                &\approx - \frac{c_P T}{l} \left( \nabla' - \delad \right) \tder{\ln P}{M_2}{m} \Mdot_2\,.
\end{split}
\end{equation} 
where the first approximation holds when mass loss effects dominate $(\partial \ln P/\partial t)_m$ and the elapsed time under consideration $\delta t \ll \tauth$.  In the outer layers, the region most heavily weighted by the $d \ln P/d M_2$ term (see Appendix \ref{app:R_response}), this last expression can be expressed approximately as
\begin{equation}
\der{\ln l}{m} \approx \frac{\tauth}{\taum}  \left( \nabla' - \delad \right)\pder{\ln P}{M_2} \,.
\label{eq:dlnldm_Mdottaus}
\end{equation}

Where $\delta t \ll \tauth \ll \taum$, heat transport is able to approximately maintain the original thermal profile, producing $\del' \approx \del$.  For radiative regions, $\del < \delad$, producing $d \ln l/dm < 0$ as mass elements must absorb flux in order to increase their entropy as they move outward.  In convective regions, $d \ln l/dm \gtrsim 0$. Since the fractional change in $l$, $\delta l/l$ due to advection goes as $\sim \tauth/\taum$, the overall flux decrement is rather small in this limit.

In the opposite limit, when $\tauth \gg \taum$, mass elements are advected outward nearly adiabatically. Thus $\del' \rightarrow \delad$ from below in radiative regions and from above in convective zones. Since $\tauth/\taum$ can become very large, significant $l$ perturbations due to mass loss can be driven by only a very slight non-adiabaticity in the advective flow.  Our numerical calculations bear this out, showing that the magnitude of $l$ perturbations are largest during the adiabatic mass loss phases.

\end{document}